\documentclass[superscriptaddress,groupedaddress,nofootnoteinbib,12pt]{article}  % for review and submission
\usepackage{graphicx}  % needed for figures
\usepackage{dcolumn}   % needed for some tables
\usepackage{bm}        % for math
\usepackage{amssymb}   % for math
\usepackage{jcappub}

\def\be{\begin{equation}}
\def\ee{\end{equation}}
\def\ba{\begin{eqnarray}}
\def\ea{\end{eqnarray}}
\def\beq{\begin{eqnarray}}
\def\eeq{\end{eqnarray}}

\def\mpl{M_{\rm Pl}}

\def\d{\mathrm{d}}
\def\p{{\cal P}}

\def\L*{{\cal L}_*}
\def\L{\mathcal{L}}
\def\({\left(}
\def\){\right)}
\def\ie{{\it i.e. }}
\def\etal{{\it et.al. }}
\def\nn{\nonumber}
\def\p{\partial}
\def\mn{_{\mu \nu}}
\def\stu{St\"uckelberg }
\def\p{\partial}

\def\<{\langle}
\def\>{\rangle}

\def\vk{\vec{k}}
\def\vy{\vec{y}}
\def\A{\mathcal{A}}
\begin{document}

\title{Chronology Protection in Galileon Models and Massive Gravity}

\author[1,2]{Clare Burrage,}
\author[1,3]{Claudia de Rham,}
\author[1]{Lavinia Heisenberg}
\author[3]{and Andrew J. Tolley}
\affiliation[1]{D\'epartment de Physique  Th\'eorique and Center for Astroparticle Physics, Universit\'e
de  Gen\`eve, 24 Quai E. Ansermet, CH-1211  Gen\`eve}
\affiliation[2]{School of Physics and Astronomy, University of Nottingham, Nottingham NG7 2RD, UK}
\affiliation[3]{Department of Physics, Case Western Reserve University, 10900 Euclid Ave, Cleveland, OH 44106, USA}
%\date{\today}

%%%%%%%%%%%%%%%%%%%%%%%%%%%%%%%%%%%%%%%%%%%%%%%%%%%%%%%%%%%%%%%%%%%%%
%%%% Abstract

%Massive Gravity in four dimensions has been shown to be free of the Boulware-Deser (BD)  ghost
%both in the ADM and the \stu languages for a  specific choice of mass terms.  We show here how this is consistent with the
%helicity language.
\abstract{Galileon models are a class of effective field theories that have recently received much attention. They arise in the decoupling limit of theories of massive gravity, and in some cases they have been treated in their own right as scalar field theories with a specific nonlinearly realized global symmetry (Galilean transformation). It is well known that in the presence of a source, these Galileon theories admit superluminal propagating solutions, implying that as quantum field theories they must admit a different notion of causality than standard local Lorentz invariant theories. We show that in these theories it is easy to construct closed timelike curves (CTCs) within the {\it naive} regime of validity of the effective field theory. However, on closer inspection we see that the CTCs could never arise since the Galileon inevitably becomes infinitely strongly coupled at the onset of the formation of a CTC. This implies an infinite amount of backreaction, first on the background for the Galileon field, signaling the break down of the effective field theory, and subsequently on the spacetime geometry, forbidding the formation of the CTC.
Furthermore the background solution required to create CTCs becomes unstable with an arbitrarily fast decay time.
Thus Galileon theories satisfy a direct analogue of Hawking's chronology protection conjecture.}

\maketitle

%%%%%%%%%%%%%%%%%%%%%%%%%%%%%%%%%%%%%%%%%%%%%%%%%%%%%%%%%%%%%%%%%%%%%
%%%% Introduction

\section{Introduction}
Mankind has long dreamed of being able to travel back in time, although the discussion of the associated causal paradoxes has an almost equally long history.  The unification of space and time in Einstein's relativity led to the possibility of considering time travel as a physical possibility that could be studied mathematically.

As human history has not been invaded by multitudes of visitors from the future, there is strong evidence that travel into the past cannot occur, however the question of whether the laws of physics allow the possibility of traveling back in time remains.
Indeed it is surprising that General Relativity (GR) has perfectly valid solutions which allow for Closed Time-like Curves (CTCs).  These are geodesic or non-geodesic curves which form closed loops in space and time which can be traversed in a time-like manner.  If a person were to travel along such a path they would arrive back at their initial point in coordinate space and time after a (positive) finite amount of proper time had elapsed.  Just as the ideas of time travel immediately raise the ideas of causal paradoxes, the existence of a CTC makes the causal structure of a spacetime impossible to determine.  There is no Cauchy surface on which initial data can be given which determines the future of the spacetime uniquely.

In 1991 Hawking argued that physics contains a mechanism which protects itself against the formation of CTCs, \cite{Hawking:1991nk}, (see Ref.~\cite{Visser:1995cc} for a comprehensive review).  The specific example he used was the two-dimensional Misner geometry however he argued that the qualitative properties of this model may hold in general. If a CTC exists in spacetime the energy-momentum of the quantum field which describes the particle trying to traverse this path becomes so large that its backreaction cannot be neglected, this is then expected to destroy the existence of the CTC by modifying the spacetime geometry itself. In modern parlance, the effective field theory of gravity inevitably breaks down before the onset of formation of the CTC. This was called Chronology Protection by Hawking.

In this paper we explore how chronology protection occurs in the Galileon model and massive gravity. One key difference worth emphasizing at the outset is that in these models there arises more than one effective metric describing the propagation of different species, specifically the effective metric seen by the Galileon, and the usual metric of the graviton. GR contains a single metric which imposes causality, and conventional matter coupled to GR is (sub)luminal, nevertheless it does admit CTCs for sufficiently peculiar geometries. In Galileon models and massive gravity, the situation is seemingly worse since on top of that they allow for fluctuations in the Galileon field which are superluminal with respect to the usual GR metric. Thus we anticipate the possible existence of new types of CTCs, not allowed in GR, which are tied to the existence of this two metric structure, \ie arise as a direct consequence of this superluminality. In fact we shall show that it is easy to find a situation whereby both metrics are well-defined in the sense that particles coupled exclusively to a single metric will never be able to form CTCs. However, allowing for the two species of particles to interact will allow for an exchange of information between the two metrics which will in turn allow for a generalization of the concept of a CTC to arise. In this generalized notion of a CTC - `bimetric CTCs' - it is sufficient that the curve is timelike only with respect to one of the two metrics at any given proper time along the curve. We shall nevertheless see that such generalized CTCs are forbidden by the same chronology protection physics.

A second crucial difference between the CTCs considered in this paper and those considered in GR is that they arise even in the decoupling limit $M_{\rm pl} \rightarrow \infty$ in which limit the graviton (GR) metric is taken to be Minkowski $g_{\mu\nu}= \eta_{\mu\nu}$. In Hawking's chronology protection it is the backreaction of the stress energy on the GR metric which is seen to forbid the CTC. However, in the present case this backreaction can be made arbitrarily small by making $M_{\rm Pl}$ arbitrarily large. Nevertheless we shall see that even before the backreaction on the metric can become significant, the effective field theory of the Galileon will breakdown because of strong coupling. In other words, the breakdown of the Galileon effective field theory (EFT) will always occur before, \ie at lower energy scales than, the equivalent breakdown in GR. This is tied to the fact that the characteristic scale of interactions in these theories $\Lambda$ is much lower than the Planck scale $\Lambda  \ll M_{\rm Pl}$.  Despite this we shall see that there is an entirely analogous (but for these reasons not identical) chronology protection mechanism in these models to the pure GR case discussed by Hawking.

The Galileon \cite{Nicolis:2008in} was proposed as a generalization of decoupling limit of the Dvali-Gabadadze-Porrati (DGP) model \cite{Dvali:2000hr},  the first theory of gravity mediated by an effectively softly massive spin-2 field without incurring ghost-like pathologies.  In the decoupling limit of DGP, \cite{Luty:2003vm}, where the Planck mass becomes infinite and the mass of the graviton vanishes in such a way that the spin-2 and spin-0 components of the massive graviton decouple, the resulting effective scalar field theory has two defining properties: The theory obeys the Galileon symmetry $\pi \rightarrow \pi + v_{\mu}x^{\mu} + c$, and the equation of motion for the scalar field is second order in derivatives despite higher order derivative terms appearing in the Lagrangian. Nicolis \etal showed that in four dimensions there are five Lagrangian operators with these two properties and they dubbed the most general scalar field theory of this form the Galileon \cite{Nicolis:2008in}.

In its original formulation the Galileon was expected to arise within the effective description of a more fundamental theory, similarly as in DGP. However it was soon considered within the community as a fundamental scalar degree of freedom in its own right, with no reference to another underlying description \cite{arXiv:0901.1314,deRham:2010eu,arXiv:1009.2497}.  Only later, was it realized that the Galileon generically appears as the helicity-0 mode in the decoupling limit of healthy theories of massive gravity both in four \cite{deRham:2010gu,deRham:2010ik,deRham:2010kj} and three dimensions \cite{deRham:2011ca}. Whether or not the graviton has a mass is another unresolved fundamental question, and the presence of a graviton mass could have important implications for Cosmology.  Observations of the solar system constrain the mass to be less than or of the order $10^{-32} \mbox{ eV}$ (this bound may change slightly depending on the two free parameters of the theory), but this could still become relevant on Hubble scale distances today and modify the evolution of the Universe.  It has been successfully shown that the construction of a massive graviton proposed in \cite{deRham:2010kj} is free of ghosts in the ADM formalism, first to fourth order \cite{deRham:2010kj} and then in full generality, \cite{Hassan:2011hr, Hassan:2011ea}. These results have been confirmed in both the \stu and helicity-languages, \cite{deRham:2011rn,deRham:2011qq}.
The Galileon also arises  as the four dimensional low energy effective field theory of certain five dimensional probe brane scenarios \cite{deRham:2010eu}, where the Galileon scalar describes the position of the brane in the fifth dimension.  Similarly they arise naturally in the decoupling limits of higher dimensional extensions of the DGP model \cite{cascading,bigalileon}. However it is the massive graviton origin of the Galileon which is of most interest for this work.

The Galileon has a broad and interesting phenomenology,  but one potentially worrying phenomenon is that fluctuations of the Galileon field can propagate superluminally \cite{Hinterbichler:2009kq,Goon:2010xh}.  Superluminal fluctuations had also previously been shown to be present in the DGP model \cite{Adams:2006sv,Hinterbichler:2009kq}.  Indeed superluminal modes are generic to Galileon \cite{Nicolis:2009qm} and massive gravity \cite{Dubovsky:2005xd,deRham:2011pt,Gruzinov:2011sq} constructions.
Superluminal fluctuations are often considered to be a symptom of a sick theory  since if a particle is traveling superluminally in one reference frame, then there exists another choice of frame in which the particle is traveling backwards in coordinate time, and from this a CTC could possibly be constructed.  However it is not guaranteed that CTCs can form \cite{Babichev:2007dw}.  In general the field that exhibits superluminal fluctuations comes with its own effective metric and causal structure which is independent to that felt by photons. Even if the causal cones of these fluctuations lie outside the causal cones of photons, the causal structure of the spacetime can be protected \cite{Babichev:2007dw} if there exists one foliation of spacetime into surfaces which can be considered as Cauchy surfaces for both metrics.

Fluctuations of the Galileon fields do carry their own effective metric and hence have an independent causal structure.  The superluminality arises when there is a non-trivial background configuration for the Galileon $\pi_0(\vec{x})$, for which fluctuations around this background configuration feel a metric $G\mn$ which depends on the metric of the background spacetime, but also on $\pi_0(\vec{x})$ and its derivatives.   For suitable choices of $\pi_0(\vec{x})$ this can allow for superluminal Galileon fluctuations.

It was suggested recently that  CTCs \cite{Evslin:2011vh}, could also be constructed in the Galileon model.  It was shown that Galileon fluctuations propagate superluminally when the background is a plane wave solution of the Galileon equations of motion.  It was then hypothesized that arranging for four of these plane waves to form a rectangle would allow for the fluctuations to travel on a CTC.  In section \ref{sec:galileon} we will consider a related scenario where the plane wave travels on a circle, and discuss the circumstances under which a CTC forms, and how the Chronology Protection Conjecture operates in this scenario.

In this paper we consider the formation of CTCs in two different setups. The first one is unique to massive gravity, whilst the second one relies on the existence of two different background metrics and could arise both in massive gravity and in more general Galileon theories. In both cases, we show explicitly that the effective metric felt by the graviton helicity-0 mode or the Galileon is identical to a well-known class of metrics of GR for which CTC may be created, and an analogue of the Hawking Chronology Protection Conjecture is found. In particular we see that when starting from healthy initial conditions, one necessarily needs to pass through a region of infinite strong coupling  to produce the required background on top of which CTCs may form. Furthermore we emphasize that modes with large enough momentum along the direction transverse to the CTC are unstable. Since one cannot prevent the excitation of such modes at the quantum level, this implies the existence of an arbitrarily fast instability.

The rest of this paper is organized as follows: We start by reviewing the formation of CTCs in GR and Hawking's Chronology Protection Conjecture. We then move to the description of the decoupling limit of massive gravity in section \ref{sec:massgrav} and as an explicit example we show how a simple gas of dust may be sufficient to produce a background solution for which CTCs may form. We show that precisely the same arguments as in GR hold in the case of massive gravity, and one can therefore extend Hawking's Chronology Protection Conjecture to massive gravity for configurations where the effective metric exhibits CTCs even if the real metric does not. We then present another class of configurations in section \ref{sec:galileon} which are applicable for both Galileon and massive gravity models. In that case CTCs may form when particles living on two different effective metrics interact. However in the regime where CTCs form, the background is unstable and decays arbitrarily fast. Furthermore, it is impossible to reach this regime without the Galileon becoming infinitively strongly coupled. This implies that the CTC could never have formed, at least within the regime of validity of the EFT. We finally conclude in section \ref{sec:Outlook} by summarizing these results and providing an outlook.

%%%%%%%%%%%%%%%%%%%%%%%%%%%%%%%%%%%%%%%%%%%%%%%%%%%%%%%%%%%%%%%%%%%%%
%%%% A review of Andrew's argument
\section{Closed Time-Like Curves in General Relativity}
\label{sec:GR}

As discussed in the introduction CTCs are  known to exist in General Relativity.   Here we review one example and explicitly demonstrate how the chronology protection occurs.
This example, although by itself somewhat over simplistic, will be closely related to our later construction of CTCs for the Galileon and massive gravity. It does however exhibit many of the generic features expected of CTCs.

We consider the cylindrical metric, \cite{Visser:1995cc}
\begin{equation}
\label{eq:GR_metric}
\d s^2 = -(\d t+\alpha(x) \d x)^2+\d x^2+\d y^2 +\d z^2\,,
\end{equation}
where $x$ is a periodic coordinate and we identify $x=x+L$.  The function $\alpha(x)$ is an arbitrary periodic function. We can always perform a coordinate transformation to replace $\alpha(x)$ with $A=  L^{-1}\int_0^{L} dx \,  \alpha(x)$ by means of $ t \rightarrow t - f(x)$ with $f'(x)= \alpha(x) - A$, however it is never possible to remove the constant $A$ because the associated coordinate transformation $f(x)$ would not be periodic.

The null vectors which point along the spatial $x$ direction are
\begin{equation}
v_1=\left(\begin{array}{c}
1-\alpha\\
1\\
0\\
0
\end{array}\right)\;,\;\;\;\;\;
v_2=\left(\begin{array}{c}
1+\alpha\\
-1\\
0\\
0
\end{array}\right)\,.
\end{equation}
It is clear that for sufficiently large $|\alpha|$ the causal cones tips over to include points at previous values of coordinate time.
Propagating backwards in coordinate time at one instant is not enough to create a CTC. However this metric can give rise to CTCs for a suitable choice of $\alpha(x)$ or more precisely $A$.

For concreteness, let us consider a ``right-mover" particle. If it starts from the origin at $t=0$, it will reach the point $\bar x>0$ after a time,
\begin{equation}
t(\bar x)=a \bar x-\int_0^x\alpha(x)\d x\,,
\end{equation}
where $a$ is a constant. This is a null geodesic if $a=1$, and timelike if $a>1$.
The geodesic path returns to the same point in space after traversing the interval $x\in [0,L]$ at time
\begin{equation}
T=a L-\int_0^L\alpha(x)\d x\,,
\end{equation}
 and this occurs at negative coordinate time if $\alpha$ is such that the following condition is satisfied
\begin{equation}
1\le a<\frac{1}{L}\int^L_0\alpha(x)\d x=A\,.
\end{equation}
This condition can only be satisfied if there exists at least one point in the interval $x\in[0,L]$ for which $\alpha(x)>1$ (this is a necessary but not sufficient condition).  For a time-like curve, increasing $a$ makes this bound tighter.

\subsection{Instabilities}

So it is possible for  CTCs to  exist in such a metric, but the Chronology Protection Conjecture teaches us that physics can protect itself from particles  traveling around these paths. To see that this is exactly what happens for this metric consider the action for a massless  scalar field on this background with action
\begin{equation}
S=\int \d^4x\; \frac{1}{2}\left[(1-\alpha^2)(\partial_t\phi)^2+2\alpha \partial_t\phi\partial_x\phi-(\partial_x\phi)^2-(\partial_y\phi)^2-(\partial_z\phi)^2\right],
\end{equation}
and equation of motion
\begin{equation}
-(1-\alpha^2)\partial_t^2\phi-2\alpha \partial_x\partial_t\phi -\alpha^{\prime}(x)\partial_t\phi+\partial_x^2\phi +\partial_y^2\phi+\partial_z^2\phi=0.
\end{equation}
The general solutions can be written as superpositions of eigensolutions of the form
\begin{equation}
\phi=\phi_0 \exp\left(-iEt+i \kappa x+i\vk . \vy -iE\int^x_0\alpha(\bar{x})\d\bar{x}\right),
\end{equation}
where from now on we use the notation $\vk=\{k_y,k_z\}$ and $\vy=\{y,z\}$. $\phi_0, E, \kappa, k_{y,x}$ are constants that satisfy the familiar mass-shell condition
\begin{equation}
E^2=\kappa^2+k^2\,,
\label{eq:kappa1}
\end{equation}
with $k^2=|\vk|^2=k_y^2+k_z^2$.
Solutions to the equation of motion must be periodic under $x\rightarrow x+L$ therefore the momentum along the $x$-direction must be quantized, and so is the energy:
\begin{equation}
2\pi n= \kappa_{n} L-E(n,\vk) \int^L_0\alpha(x) \d x = \kappa_n L - E(n,\vk) A\, L\,, \hspace{20pt}{\rm with} \hspace{10pt}n \in \mathbb{Z}\,.
\label{eq:kappa2}
\end{equation}
Combining (\ref{eq:kappa1}) and (\ref{eq:kappa2}), % and writing $A=(1/L)\int^L_0\alpha d\bar{u}$
gives an equation for the quantized allowed energies $E_n$
\begin{equation}
\label{energies}
E_{\varsigma}(n,\vk)=\frac{1}{(1-A^2)}\left[\frac{2\pi n}{L} A +\varsigma \sqrt{\(\frac{2\pi n}{L}\)^2+ (1-A^2)k^2 }\ \right].
\end{equation}
When the field is quantized, we should split the modes up into positive and negative energy solutions. In this case, this corresponds to a different choice of the sign in front of the square root. These satisfy $E_{\varsigma=-1}(-n)=-E_{\varsigma=+1}(n)$. From now on we focus on the positive frequency mode only ($\varsigma=+1$) and drop the subscript for simplicity (\ie $E\equiv E_{\varsigma=+1}$).
Putting this together, the correctly normalization quantized fields are
\ba
\phi(t,x,\vy) = \frac{1}{L}\sum_{n=-\infty}^{\infty} \int \frac{\d^2 \vk}{(2 \pi)^2} \left( a_{n,\vk}\, u_{n,\vk}(x,\vy,t)
+  a^{\dagger}_{n,\vk} u^*_{n,\vk}(x,\vy,t)  \right),
\ea
where
\be
[a_{n,\vk},a^{\dagger}_{n',\vk'}] = (2\pi)^2\, L \, \delta_{n n'}\, \delta^{(2)}(\vk-\vk')\,,
\ee
and
\be
u_{n,\vk}(x,\vy,t) = \frac{1}{\sqrt{2(E-A \kappa_n)}} \exp\left(-iE t+i \kappa_n x+i\vk . \vy -iE\int^x_0\alpha(\bar{x})\d\bar{x}\right).
\ee
The unfamiliar factor
\ba
\sqrt{(E-A \kappa_n)}= \(\frac{4\pi^2n^2}{L^2}+k^2(1-A^2)\)^{1/4}
\ea
arises from the Klein-Gordon inner product normalization.

Clearly when $|A|>1$ the energy becomes complex for fixed $n$ at sufficiently high $k$, meaning both that the solutions become unstable, and that the theory cannot be quantized. Interesting had we concentrated on a purely two dimensional model of the CTC (\ie $\vk=0$), it would not have been possible to see this instability. We shall see a similar result in the CTCs we find later. Furthermore, the instability gets faster at larger $k$ meaning that the instability is always faster than any other dynamical time scale in the system.

In the above we have quantized the fields following the standard canonical procedure applied to surfaces of constant $t$. This is the correct procedure for $|A|<1$ for which $t$ is a good time coordinate. However for $|A|>1$ surfaces of constant $t$ are no-longer spacelike, which is precisely the condition necessary to allow for the formation of CTC. Nevertheless, what the above solutions show is that regardless of how we choose to quantize in that region, an instability is present since for $|A|>1$ eq.~\eqref{energies} implies either that $E$ is complex at sufficiently large $k$ or that $k_y$ or $k_z$ are complex. Crucially it is the quantization condition which allows us to conclude this, and in this sense this is a purely quantum mechanical effect which would not have been seen by simply looking at the classical geodesics.\footnote{Of course the same effect is seen for classical fields, but the interpretation of fields in terms of particles is intrinsically quantum.}

An obvious criticism of solutions of this type that has been often levied is that if $|A|>1$ and $A$ is independent of time, then the CTC has existed for all time,  this is then an example of what has been called the `garbage in - garbage out' principle, \cite{Visser:1995cc}. We should only really worry about CTCs if they can be shown to inevitably arise from well-defined initial data. However, the key point is that even if we imagine forming this CTC from some previously well defined solution, for instance let us imagine $A$ evolves in time from the region $|A|<1$ to the region $|A|>1$, no matter how rapidly the CTC forms, there are always quantized modes of $\phi$ in the full theory which have sufficiently large momenta that they vary more rapidly than the function $A(t)$. For these modes, it is always possible to perform a WKB approximation, for which the approximate WKB energies are as in \eqref{energies} with $A$ now replaced by the appropriate function of $t$.
Starting from $|A| <1$, the energy of the solution becomes infinite when $A^2=1$. This is exactly the condition that was required for the existence of a CTC.
At this point the backreaction of the quantum fluctuations of the scalar on the background geometry will become infinitely large. This can be seen explicitly by either calculating the expectation value of the $n$-point functions of the scalar or for instance by calculating the Casimir energy associated with the compact direction $x$. Therefore we clearly see how the Chronology Protection Conjecture manifests in this scenario: If the CTC is assumed to form then particles become unstable (or more precisely the notion of particles is ill-defined) and we can no longer trust the EFT description of the scalar coupled to gravity. However, before the onset of the formation of the CTC the EFT will become strongly coupled due to the infinite energies associated with arbitrarily large but finite momenta modes.

\subsection{Strong Coupling and Quantum Backreaction}
\label{sec:strong1}

The essence of the chronology protection mechanism is that quantum fluctuations become arbitrarily large before the onset of the formation of the CTC, preventing either the formation, or at least its description within the EFT.
To understand how the chronology protection works consider the Hamiltonian for the scalar. The conjugate momentum for the scalar field is
\ba
p_\phi=(1-\alpha^2)\partial_t \phi +\alpha \partial_x \phi\,,
\ea
making the Hamiltonian
\ba
\mathcal{H}&=&\frac{\( p_\phi -\alpha \partial_x\phi\)^2}{2(1-\alpha^2)}+\frac{1}{2} \((\partial_x\phi)^2+(\partial_y\phi)^2+(\partial_z\phi)^2 \)\\
&=&\frac{1}{2}(1-\alpha^2) (\partial_t \phi)^2+\frac{1}{2} \((\partial_x\phi)^2+(\partial_y\phi)^2+(\partial_z\phi)^2 \)\,.
\ea
For fixed conjugate momentum $p_{\phi}$ whenever a solution crosses $\alpha=1$, which we recall was a necessary condition for the existence of a CTC, the Hamiltonian diverges unless simultaneously $p_{\phi}=\partial_x \phi$ at that point. Such a condition would remove one of the phase space degrees of freedom, and is invariably inconsistent at the quantum level as we shall now show.

First let us compute the quantum expectation value $\langle 0 | \phi^2(x) | 0 \rangle $. As usual, this is infinite, however it contains an $L$ dependent contribution which is finite due to the fact that the divergent terms can all be removed with local counterterms which are $L$ independent. Plugging in the quantum modes, and computing in the standard vacuum that satisfies $a_{n,\vk}|0 \rangle =0$ we find
\be
\langle 0 | \phi^2(x) | 0 \rangle =  \frac{1}{L}\sum_{n=-\infty}^{\infty} \int \frac{\d^2 \vk}{(2 \pi)^2} \frac{1}{2 \sqrt{(2 \pi n/L)^2+k^2 (1-A^2)}}\,.
\ee
To compute this let us use zeta function regularization (including the parameters $\mu$ and $s\to0$) and replace with
\ba
\langle 0 | \phi^2(x) | 0 \rangle =  \lim_{s\to 0}\, \frac{1}{L}\sum_{n=-\infty}^{\infty} \int \frac{\d^2 \vk}{(2 \pi)^2} \frac{1}{2} \mu^s ((2 \pi n/L)^2+k^2 (1-A^2))^{-1/2-s}\, .
\ea
The integral over $\vk$ may easily be performed by going to polar coordinates in the $y-z$ plane to give
\ba
\langle 0 | \phi^2(x) | 0 \rangle =  \lim_{s\to 0}\frac{1}{2 \pi(1-2s)} \frac{1}{L(1-A^2)}
\sum_{n=-\infty}^{\infty} \frac{1}{2} \mu^s \left| \frac{2 \pi n}{L}\right|^{1-2s} \,.
\ea
Now in the limit $s \rightarrow 0$ and using the fact that the Riemann zeta function satisfies $\zeta(-1)=-1/12$ so that
$\sum_{n=-\infty}^{\infty} |n| = 2 \zeta(-1)=- 1/6$ we obtain
\be
\langle 0 | \phi^2(x) | 0 \rangle = -\frac{1}{12} \frac{1}{L^2(1-A^2)}.
\ee
Having removed the quadratic divergence, we should really think of this as a finite contribution to this correlation function\footnote{In the absence of gravity we could be easily persuaded that this contribution could also be absorbed via a renormalization. However, $L$ is physically a nonlocal function of the metric, and just as in the usual Casimir effect, it has a physical consequence due to the fact that $L$ can change dynamically.}. Its importance is that it will show up in interactions, for instance, in the mean field approximation we would replace a $\frac{1}{4!}\lambda \phi^4$ interaction with a term of the form $\frac{1}{2} \lambda \langle 0 | \phi^2(x) | 0 \rangle \phi^2$. More generally it contributes to the Feynman propagator any so this $L$ dependence shows up in any perturbative calculation. The crucial point is that as $|A|$ approaches unity from below, the two point correlation function diverges. Since this term is $L$ dependent it cannot be absorbed by a local counterterm. The fact that the $\phi$ correlation function diverges even after renormalization indicates that any $\phi$ self interactions will become infinitely large indicating strong coupling of the $\phi$ field. However, even if this field is assumed to be exactly free from self-interactions, we cannot switch off the interactions of it with gravity. Thus the backreaction of it on the geometry is similarly divergent. To see this let us calculate the expectation value of the Hamiltonian. This serves as an indicator of the magnitude of the backreaction of the scalar on the spacetime geometry. We have
\ba
\langle 0 | \mathcal{H} | 0 \rangle = \frac{1}{L}\sum_{n=-\infty}^{\infty} \int \frac{\d^2 \vk}{(2 \pi)^2}  \frac{1}{4(E-A \kappa_n)} \( (1-A^2)E^2 + \(\frac{2 \pi n}{L}\)^2+ k^2 \)\,.
\ea
This expression is of course again infinite before renormalization, but contains a finite $L$-dependent contribution which accounts for the Casimir effect in this geometry. By performing a zeta function regularization, we may replace this expression with
\ba
\langle 0 | \mathcal{H} | 0 \rangle =  \lim_{s\to 0} \frac{\mu^s }{L}\sum_{n=-\infty}^{\infty} \int && \!\! \frac{\d^2 \vk}{(2 \pi)^2}
\frac{1}{4}\(\(\frac{2 \pi n}{L}\)^2+ (1-A^2)k^2\)^{-(1+s)} \\
&& \times \( (1-A^2)E^2 + \(\frac{2 \pi n}{L}\)^2+ k^2\)\,.\nn
\ea
As before, first performing the integral over $k$ gives
\ba
\langle 0 | \mathcal{H} | 0 \rangle =  \lim_{s\to 0}
\frac{\mu^s }{8 \pi L (1-A^2)^2}\(\frac{2}{3-2s}-\frac{A}{s-1}\)\sum_{n=-\infty}^{\infty} \left| \frac{2\pi n}{L}\right|^{3-2s}\,,
\ea
which in the limit using $\sum_{-\infty}^\infty |n|^3=2\zeta(-3)=1/60$ gives
\ba
\langle 0 | \mathcal{H} | 0 \rangle =
\frac{\pi^2}{60 }\(\frac{2}{3}+A\)\frac{1}{L^4 (1-A^2)^2}\,.
\ea
As in the case of the two-point function, this energy density becomes infinite at the onset of formation of the CTC. Once the associated curvature $R \sim \langle 0 | \mathcal{H} | 0 \rangle /M_{\rm Pl}^2$ becomes of order $M_{\rm Pl}^2$ the effective field theory of gravity breaks down. This clearly occurs when $L \sqrt{1-A^2}$ is of order $M_{\rm Pl}$. Thus in practice this amounts to saying that when the proper length of the closed loop is Planckian, we can no longer trust the background solution that gives rise to the CTC.

The metric \eqref{eq:GR_metric} is, after an appropriate coordinate redefinition, precisely the same as that presented in Ref.~\cite{Visser:1995cc} in the toy model of section 19.3.2, with $t^{\rm there}=t+\int \alpha(x)\d x$, such that the identification $(t,0)\equiv (t,L)$ here corresponds to the identification
$(t^{\rm there},0)\equiv (t^{\rm there}+L A,L)$ there. As defined in eq.~(19.9) of \cite{Visser:1995cc},  the `once-through-the-wormhole' interval is then
\ba
\label{otw_interval}
s=L\sqrt{1-A^2}\,,
\ea
This interval measures the proper length around the closed curve. When $s^2>0$ the curve is spacelike, when $s^2<0$ is is timelike. We thus see that the two-point function of the scalar fluctuations scale as $1/s^2$ and the energy density scales as $1/s^4$.

As discussed in  \cite{Visser:1995cc} the essence of the chronology protection is that generically CTCs, which are assumed to arise at some finite time, arise from a previously closed spacelike curve becoming timelike. Continuity guarantees that this can only occur if at the transition point the curve is null, in which case the proper distance around it is zero. This is the distance captured by $s=L\sqrt{1-A^2}$. Quantum mechanics, embodied in the above Bohr-Sommerfeld type quantization condition guarantees that there are quantum contributions to correlation functions and energy densities which scale as positive powers of $1/s$, at the scale anticipated by the uncertainty principle. At the onset of the formation of the CTC we have $s=0$ and so all these non-local contributions to correlation functions diverge. In practice this means there must be some finite positive $s$ at which point the effective field theory in question has broken down. In the present case this is the effective field theory of gravity, and the fact that it is breaking down is often stated as a breakdown of the semi-classical expansion. \\

In short, chronology protection will occur whenever the following two criterion are met:

\begin{itemize}

\item The formation of the CTC is associated with the proper distance $s$ around a closed loop transitioning from spacelike to timelike.

\item Quantum effects, associated with fluctuations around the loop scale as positive powers of $1/s$.

\end{itemize}

In addition, if the CTC is assumed to form, fields become unstable at an arbitrarily fast rate, at least in dimensions $d>2$. To prove the chronology protection conjecture we would have to prove that the previous two conditions are always met. Unfortunately this is a difficult task and so the conjecture has so far been mainly confronted on a case by case basis. Nevertheless, in the cases where the CTCs are at least approximately geodesics, a condition which does not have to hold, then we always expect an analogue of the Bohr-Sommerfeld quantization rule to apply, which appears to be sufficient to guarantee the conjecture.

\subsection{Localized CTCs}

The previous example, is clearly very special since it was necessary to assume that the topology of the Universe was compact in the $x$ direction, and the CTCs were associated with traveling around this compact direction. A more physically interesting example is a CTC which is localized in space in an otherwise asymptotically well-defined spacetime. This is easy to achieve, the previous example is a natural limit of a well-known model with localized CTCs, namely the spinning cosmic string metric:
\be
\d s^2 = -\(\d t +  J \d\theta\)^2 + \d r^2 +\(1-\delta\)^2  r^2 \d \theta^2 + \d z^2\,,
\ee
where $2\pi \delta$ is the deficit angle, $\theta = [0,2\pi]$ and $J$ is proportional to the angular momentum of the string. In the region of fixed $r$ this metric looks formally the same as the previous example. However, now the $r$ dependence allows for the CTCs to be spatially localized. A straightforward calculation shows that curves at fixed $r$ and $z$ are CTCs when
\be
r < \frac{J}{1-\delta}\,.
\ee
In practice the string has some finite core width $L_W$ and the above metric is only valid for $r>L_W$ and CTCs can only be said to exist if $J>L_W (1-\delta)$. As in the previous case, we can imagine forming the CTC at finite time, by adiabatically changing the value of $\delta$ or $L_W$ or $J$ so that there is a transition from the regime for which $J<L_W(1-\delta)$ for which CTCs are expected to be absent, to the case $J>L_W (1-\delta)$ when they will be sure to exist at distances $r < \frac{J}{1-\delta}$. In this sense it should be possible to create CTCs which are localized in space and in turn have not existed for all times (\ie are not eternal time-machines). To understand if chronology protection is applicable in this case it would be necessary to quantize the fluctuations about the string taking into account what happens inside the core of the string since that is crucial to provide the transition barrier $J>L_W(1-\delta)$. This is beyond the scope of the present work. There exist several other variations on this theme in the literature, notably the Gott and Grant solutions, see \cite{Visser:1995cc} for an extensive review.

%%%%%%%%%%%%%%%%%%%%%%%%%%%%%%%%%%%%%%%%%%%%%%%%%%%%%%%%%%%%%%%%%%%%%
%%%% Massive gravity review
\section{Galileons as Helicity-0 Modes of Massive Spin-2 Fields}
\label{sec:massgrav}
\subsection{Decoupling limit of Ghost-free Massive Gravity}

Although the idea of the Galileon arose out of looking at the decoupling limits of higher dimensional braneworld models such as the DGP model \cite{Dvali:2000hr,Luty:2003vm} and Cascading Gravity \cite{cascading}, it is in the context of four dimensional massive gravity models where the full connection between Galileons and infrared modified gravity has been easiest to explore \cite{deRham:2010gu,deRham:2010ik,deRham:2010kj}.
The first fully consistent theory of massive gravity in four dimensions was proposed in \cite{deRham:2010kj}.  The degrees of freedom of a massive spin two field can be arranged into a massless helicity-2 field, two helicity-1 modes and a helicity-0 field.  To see this explicitly, one can include four \stu fields $\phi^a$, and consider non-derivative interactions for the tensor $H\mn=g\mn-\eta_{ab}\p_\mu \phi^a\p_\nu \phi^b$.
For a graviton of mass $m$, the Lagrangian is then
\begin{equation}
\mathcal{L}=\frac{\mpl^2}{2}\sqrt{-g}\left(R-\frac{m^2}{4}\mathcal{U}(g,H)\right)\,.
\end{equation}
Then defining $\mathcal{K}^{\mu}_{\nu}(g,H)=\delta^{\mu}_{\nu}-\sqrt{\delta^{\mu}_{\nu}-H^{\mu}_{\nu}}$ the most general potential $\mathcal{U}$  that has no ghosts is, \cite{deRham:2010kj}
\begin{equation}
\mathcal{U}(g,H)=-4(\mathcal{U}_2+\alpha_3\mathcal{U}_3+\alpha_4\mathcal{U}_4),
\end{equation}
where the $\alpha_n$ are free parameters, and
\begin{eqnarray}
\mathcal{U}_2&=& [\mathcal{K}]^2-[\mathcal{K}^2],\\
\mathcal{U}_3&=& [\mathcal{K}]^3-3[\mathcal{K}][\mathcal{K}^2]+2[\mathcal{K}^3],\\
\mathcal{U}_4&=& [\mathcal{K}]^4-6[\mathcal{K}^2][\mathcal{K}]^2+8[\mathcal{K}^3][\mathcal{K}]+3[\mathcal{K}^2]^2-6[\mathcal{K}^4]\,,
\end{eqnarray}
where $[\ldots]$ represents the trace of a tensor with respect to the metric $g\mn$.
 The absence of ghost for this theory has been shown in the decoupling limit in \cite{deRham:2010gu,deRham:2010ik,deRham:2010kj}, fully non-linearly beyond the decoupling limit in \cite{Hassan:2011hr,Hassan:2011ea}, as well as in the \stu and helicity languages in \cite{deRham:2011rn,deRham:2011qq}.

In the decoupling limit, where the Planck mass is taken to infinity $\mpl \rightarrow \infty$ and the graviton mass tends to zero $m\rightarrow 0$ while the strong coupling scale $\Lambda^3 =\mpl m^2$ stays fixed, the helicity-1 mode decouples and can consistently be set to zero, and the scalar mode takes on a Galileon form, \cite{deRham:2010ik}. The tensor $H\mn$ is then of the form
\ba
H_{\mu\nu}=\frac{1}{\mpl}\(h_{\mu\nu}+\frac{2}{\Lambda^3}\Pi_{\mu\nu}-\frac{1}{\Lambda^6}\eta^{\alpha\beta}\Pi_{\mu\alpha}\Pi_{\nu\beta}\)\,,
\ea
with $\Pi_{\mu\nu}=\p_{\mu}\p_{\nu}\pi$.  $\pi$ is the canonically normalized helicity-0 component of the massive graviton and $h_{\mu\nu}=\mpl (g\mn-\eta\mn)$.
In the decoupling limit the Lagrangian reduces to
\begin{equation}
\mathcal{L}=-\frac{1}{2}h^{\mu\nu}\mathcal{E}^{\alpha\beta}\mn h_{\alpha\beta}+\sum_{n=1}^3\frac{a_n}{\Lambda^{3(n-1)}} h^{\mu\nu}X^{(n)}_{\mu\nu}
+\frac{1}{2\mpl}h^{\mu\nu}T_{\mu\nu},
\end{equation}
where  $\mathcal{E}^{\alpha\beta}\mn$ is the Lichnerowicz operator. External sources $T\mn$ scale in this decoupling limit in such a way that the quantity $\tilde T\mn = T\mn/\mpl$ remains finite.
The coefficients $a_n$ are related to the constants $\alpha_n$, with $a_1=1$ and
\begin{eqnarray}
X^{(1)}_{\mu\nu}&=& \Box \pi g_{\mu\nu}-\Pi_{\mu\nu}\\
X^{(2)}_{\mu\nu}&=& \Pi_{\mu\nu}^2-\Box\pi\Pi_{\mu\nu}-\frac{1}{2}([\Pi^2]-[\Pi]^2)g_{\mu\nu}\\
X^{(3)}_{\mu\nu}&=& 6\Pi^3_{\mu\nu}-6[\Pi]\Pi^2_{\mu\nu}+3([\Pi]^2-[\Pi^2])\Pi_{\mu\nu}-g_{\mu\nu}([\Pi]^3-3[\Pi^2][\Pi]+2[\Pi^3])\,,
\end{eqnarray}
% The ghost free property means that these terms are closely related to the Galileon operators and obey the same symmetries.
where   $\Box\pi=\partial_{\mu}\partial^{\mu}\pi$.

For simplicity we will study a massive gravity theory with $a_3=0$, and  then absorb the value of $a_2$ into the scale $\Lambda$ and set $a_2=1$.
The fields can then be redefined to diagonalize the kinetic terms of the scalar and spin two degrees of freedom giving
\ba
\mathcal{L}=-\frac{1}{2}h^{\mu\nu}\mathcal{E}^{\alpha\beta}\mn h_{\alpha\beta}+\frac{3}{2}\pi\Box\pi-\frac{3}{2\Lambda^3}(\partial\pi)^2\Box\pi
+\frac{1}{2\Lambda^6}(\partial\pi)^2([\Pi^2]-[\Pi]^2)\nn \\
+\pi \tilde T-\frac{1}{\Lambda^3}\partial_{\mu}\pi\partial_{\nu}\pi \tilde T^{\mu\nu}\,.
\ea
The equation of motion for $\pi$ is then
\ba
\label{eq:pi_massive}
3 \Box \pi- \frac{3}{\Lambda^3}\([\Pi^2]-[\Pi]^2\)+\frac{1}{\Lambda^6}\([\Pi]^3-3 [\Pi][\Pi^2]+2 [\Pi^3]\)=-\tilde T +\frac{2}{\Lambda^3}\p_\mu\p_\nu \pi \tilde T^{\mu\nu}. \hspace{15pt}
\ea
As already pointed out in Refs.~\cite{deRham:2010ik,arXiv:1010.1780,arXiv:1101.1295}, the additional coupling to matter of the form $\p_\mu\pi \p_\nu \pi T^{\mu\nu}$ can play a crucial role and distinguishes models of this type from generalized Brans-Dicke models where the Brans-Dicke scalar only couples to the trace of the stress energy. In what follows we consider a specific configuration which allows not only for the superluminal propagation of $\pi$ fluctuations but as well for the generation of CTCs.

\subsection{Closed Time-Like Curves}

Similarly to the pure GR case, we consider a cylindrical background, where the coordinate $x$ is periodic  so that $x\rightarrow x+L$, however unlike the previous section, we can here focus on a completely flat geometry $g\mn = \eta\mn$. The appearance of CTCs is then due to the fact that $\pi$ fluctuations live on a different effective metric $G\mn$ as we shall see in what follows.
To create a non-trivial background for the field $\pi$, we consider a perfectly innocent fluid of  (pressureless) dust with energy density $\rho$
\ba
\tilde \rho= \frac{\rho}{\mpl}=\frac{3}{4}\alpha^2 \Lambda^3\,,
\ea
where $\alpha$ is an arbitrary dimensionless constant. This dust will source the metric perturbation, however the metric can still be consistently treated as flat at distance scales well below the curvature scale which for this solution is set by $1/m$ where $m$ is the graviton mass (assuming $\alpha$ is of order unity). Thus as long as the size of the compact direction $L$ is significantly smaller than $1/m$, \ie $mL\gg 1$ we may consistently ignore the effect of the metric.

On the other hand, this source creates a non-trivial configuration $\pi_0(x,t)$ for the helicity-0 mode obtained by solving \eqref{eq:pi_massive}
\ba
\pi_0(t,x)= - \frac{\Lambda^3}{8}(\alpha^2 x^2+4 \alpha t x)\,.
\label{eq:pimassgrav}
\ea
A point of concern is that the solution $\pi_0$ does not appear periodic in $x$. However in massive gravity  $\pi$ is not a fundamental object, this role is instead played by $\p_\mu\p_\nu \pi$ which {\it is} periodic in the coordinate $x$ for this solution. Stated differently, under a coordinate shift $x \rightarrow x+L$ the field transforms as $\pi \rightarrow \pi - \frac{\Lambda^3\alpha L}{4}(\alpha x + 2 t)-\frac{\Lambda^3 L^2}{8}$ which is simply a Galileon transformation. In massive gravity, the system is exactly invariant under this transformation and hence the physics is unchanged.
This solution is unbounded at infinity but again this is not a physical problem since only $\p_\mu\p_\nu \pi$, which is constant, is meaningful. Furthermore since we only consider this configuration in a small enough region of spacetime $x^2, t^2 \ll \mpl / \Lambda^3=m^{-2}$ the background for $\pi$ is under control in this region.

We now consider fluctuations around this configuration, $\pi=\pi_0+\phi$, which have the following Lagrangian,
\ba
\mathcal{L}=-\frac 32 (G^{-1})^{\mu\nu}\p_\nu \phi \p_\nu \phi\,,
\ea
with the effective metric $G\mn$ satisfying,
\begin{equation}
(G^{-1})^{\mu\nu}=\eta^{\mu\nu}-\frac{2}{\Lambda^3}(K^{\mu\nu}-K\eta^{\mu\nu})
+\frac{2}{\Lambda^6}\left(K^{\mu\alpha}K_{\alpha}^{\nu}-KK^{\mu\nu}-\frac{1}{2}\([K^2]-[K]^2\)\eta^{\mu\nu}\right) +\frac{2}{3\Lambda^3}T^{\mu\nu}
\end{equation}
where $K\mn=\partial_{\mu}\partial_{\nu}\pi_0$. Note that somewhat uniquely to massive gravity models, the effective metric is explicitly a function of the local stress energy as a consequence of the $\p_\mu\p_\nu \pi \tilde T^{\mu\nu}$ in the action.

For the present background configuration the inverse effective metric felt by the fluctuations is
\ba
(G^{-1})^{\mu\nu}=\(
\begin{array}{cccc}
-1+\alpha^2 & -\alpha & 0 & 0 \\
-\alpha & 1 & 0 & 0 \\
 0 & 0 & 1 & 0 \\
 0 & 0 & 0 & 1
\end{array}\)\,,
\ea
so the effective metric $G\mn$ is identical to that discussed in \eqref{eq:GR_metric} in the case of GR.
Therefore, completely analogously to section \ref{sec:GR}, CTCs can exist in the metric felt by the fluctuations of the helicity-zero mode. Unlike the situation in GR however, this happens whilst the true spacetime metric $g_{\mu\nu}$ remains flat. Nevertheless, the same argument goes through and as discussed in section \ref{sec:GR}, to construct such a curve requires that $|A|=|1/L \int^L_0 \alpha \d x |> 1$, which in turn implies that the solution itself is unstable, with arbitrarily large instability scale $k$.

Furthermore, if one were to start with a healthy configuration for which $\alpha^2<1$ initially then no CTCs would be present in the initial setup. We could then raise the question of whether one could adiabatically change the settings so as to reach a regime where CTCs could be produced.
For this to happen, we need to increase the energy density $\rho$ adiabatically, or in other words make $\alpha$ a slowly time-varying parameter, and include a flux of energy for instance $T_{01}(x)$ such as to locally form a lump of matter with $\alpha$ locally increasing in time. CTCs can then arise when $\alpha$ crosses the threshold  $\alpha=1$, but as we have seen in the previous section, such a background configuration cannot be constructed without the renormalized two-point function of fluctuations around the background diverging.  Thus if we are to start from a healthy configuration, one cannot produce a background that allows for CTCs without going through a infinitively strongly coupled regime in the process.
Therefore we conclude that it would not be possible to construct such CTCs in a causal manner.

\subsection{Galileons as Goldstone/\stu fields versus Fundamental Fields}

In the next section, we move onto more generic Galileon models, where we consider $\pi$ as a fundamental degree of freedom. The key distinction in this description is that the shift and Galileon symmetry are then accidental rather than fundamental symmetries of a \stu field/Goldstone mode as was the case in the previous section. So far the fact that only the quantity $\p_\mu\p_\nu \pi$ was physical  implied that only that quantity had to satisfy the periodicity condition.  However when considering a setup where the Galileon is treated as its own fundamental scalar field, this no longer holds and one needs to impose the periodicity condition on $\pi$ as well. The previous configuration can therefore not be applied for a Galileon when it does not necessarily play the role of a Goldstone boson as in Massive Gravity, and as we shall see below one needs to work slightly harder to find a configuration that allows CTCs.

%%%%%%%%%%%%%%%%%%%%%%%%%%%%%%%%%%%%%%%%%%%%%%%%%%%%%%%%%%%%%%%%%%%%%
%%%% Galileon introduction
\section{Galileons as Fundamental Fields}
\label{sec:galileon}

On a flat spacetime the Galileon \cite{Nicolis:2008in} has the following Lagrangian
\begin{equation}
\mathcal{L}=-\frac{1}{2}(\partial \pi)^2 - \frac{c_3}{2 \Lambda^3} \Box \pi (\partial \pi)^2 +\frac{c_4}{\Lambda^6}\mathcal{L}_4(\pi)+\frac{c_5}{\Lambda^9}\mathcal{L}_5(\pi)+\pi T\;.
\label{eq:lag}
\end{equation}
The terms $\mathcal{L}_4(\pi)$ and $\mathcal{L}_5(\pi)$ are given by
   \begin{eqnarray}
   \mathcal{L}_4(\pi)&=& (\partial \pi)^2 \([\Pi]^2-[\Pi^2]\)\;,\\
   \mathcal{L}_5(\pi)&=&(\partial \pi)^2 \([\Pi]^3-3[\Pi][\Pi^2]+2[\Pi^3]\)\;.
      \end{eqnarray}
    The $c_n$ are arbitrary dimensionless coefficients which are expected to be of order one. The purely scalar part of the Lagrangian respects the Galileon symmetry, $\pi\rightarrow \pi +c+b_{\mu} x^{\mu}$ and is  defined up to total derivative terms which are irrelevant in flat space.

The Galileon operators are non-renormalisable, and so the Galileon model should be treated as an effective field theory, valid up to some scale $\Lambda$.   However the specific operators $\mathcal{L}_n$ are not renormalised by loop corrections \cite{Luty:2003vm,Nicolis:2004qq,Nicolis:2008in}.
 % The scale $\Lambda$ is expected to be related to the scale controlling the $c_n$ coefficients in the Lagrangian.  Therefore from this point onwards we write them as $\tilde{c}_n=c_n/\Lambda^{2(n-3)}$, where the $c_i$ are now dimensionless constants, expected to be of order one.
 The classical contribution of the  $(n+1)$-th Galileon operator is suppressed compared to the $n$-th operator by a scale which can be written schematically as
\begin{equation}
\alpha_{\rm Cl} \equiv \frac{\partial^2\pi}{\Lambda^3}\,.
\end{equation}
This is the dimensionless measure of when the non-linearity of the Galileon kinetic terms becomes relevant for classical calculations.  There is another dimensionless parameter in the theory, which schematically is
\begin{equation}
\alpha_{\rm Q} \equiv \frac{\partial^2}{\Lambda^2}
\end{equation}
which is the parameter suppressing quantum loop corrections \cite{Nicolis:2004qq,Nicolis:2008in,arXiv:1002.4873}.  The two parameters $\alpha_{\rm Cl}$ and $\alpha_{\rm Q}$ are distinct, and this allows for solutions for which the non-linearities of the Galileon become important classically $\alpha_{\rm Cl}\sim 1$, whilst quantum corrections are still well under control $\alpha_{\rm Q} \ll 1$  \cite{Nicolis:2004qq,arXiv:1002.4873}.  Thus non-linear terms can be important classically without the breakdown of the effective field theory.

The equation of motion for the scalar field obtained from the Lagrangian (\ref{eq:lag}) is
\begin{eqnarray}
\label{eq:eom}
&& [\Pi]+\frac{c_3}{\Lambda^3}\([\Pi]^2-[\Pi^2]\)
-\frac{c_4}{\Lambda^6}\([\Pi]^3-3[\Pi][\Pi^2]+2[\Pi^3]\) \\
-&&\frac{c_5}{\Lambda^9}\([\Pi]^4-6[\Pi]^2[\Pi^2]+8[\Pi][\Pi^3]
+3[\Pi^2]^2-6[\Pi^4]\)=-T\nonumber\;.
\end{eqnarray}
For the solutions we consider the higher order Galileon interactions will all vanish and so it is sufficient to study the simplest version of the Galileon theory with $c_4=c_5=0$. We can therefore absorb the coefficient $c_3$ into $\Lambda$, so as to set $c_3=1$.  Assuming once again a flat cylindrical background metric, where the $x$ coordinate is periodic under the shift $x\rightarrow x+L$ (with $L$ positive), and some background configuration $\pi_0(\vec{x})$, fluctuations $\phi$ about this background are described by the Lagrangian
\begin{equation}
\mathcal{L}_{\phi}=-\frac{1}{2}\(G^{-1}\)^{\mu\nu}\partial_{\mu}\phi\partial_{\nu}\phi-\frac{1}{2\Lambda^3}\Box\phi(\partial \phi)^2\,.
\end{equation}
Here again, the causal structure that determines the propagation of fluctuations about this background is set not by the background metric $g_{\mu\nu}$ but by the tensor
\begin{equation}
\(G^{-1}\)^{\mu\nu}=g^{\mu\nu}-\frac{2}{\Lambda^3}(K^{\mu\nu}-K g^{\mu\nu})\,,
\label{eq:galileonmetric}
\end{equation}
where as previously, $K\mn=\p_\mu\p_\nu \pi_0$.
We will see that under the right circumstances this metric $G\mn$ combined with the real spacetime metric $g\mn$ allows for the formation of CTCs.

%%%%%%%%%%%%%%%%%%%%%%%%%%%%%%%%%%%%%%%%%%%%%%%%%%%%%%%%%%%%%%%%%%%%%
\subsection{Galileon Fluctuations}

If the Galileon is treated as an effective field theory with no knowledge of its UV completion, then the  solution for $\pi$ which gives rise to CTCs in the massive gravity scenario (\ref{eq:pimassgrav}) is not acceptable because $\pi$ is not periodic in the coordinate $x$. Imposing that solutions for $\pi$ must be periodic in $x$ makes it more difficult to construct a CTC in the Galileon model.
The metric controlling the propagation of Galileon fluctuations in the simplest scenario with $c_4=c_5=0$ was given in Equation (\ref{eq:galileonmetric}). Unfortunately there is no solution for periodic $\pi_0$ for which $G_{\mu\nu}$ matches the periodic metric we discussed before in section \ref{sec:GR}, so we have to try a little harder to construct a CTC.

In Ref.~\cite{Evslin:2011vh} it was suggested that a CTC could be created in a Galileon theory on a background of plane wave solutions.  As plane wave solutions satisfy $\Box\pi_0=0$, this  simplifies the form of $G^{\mu\nu}$.  Solutions to the wave equation on a flat Cartesian background can have the form
\begin{equation}
\label{eq:pi0}
\pi_0=\frac{\Lambda^3}2 f(x+t)\,,
\end{equation}
for an arbitrary periodic function $f(y)=f(y+L)$, which is such that $f^{\prime\prime}$ is dimensionless.
The metric $G\mn$ felt by fluctuations around this background solution is then given by
%\begin{equation}
%\(G^{\mu\nu}=\left(\begin{array}{cccc}
%-1-\beta & \beta & 0 & 0 \\
%\beta & 1+\beta & 0 & 0 \\
%0& 0 & 1 & 0 \\
%0& 0 & 0 & 1
%\end{array}\right)\,,
%\end{equation}
\ba
\label{eq:eff_metric}
G\mn \d x^\mu \d x^\nu=-\d t^2 +\d x^2 + f^{\prime\prime} \(\d t+\d x\)^2+\d y^2 +\d z^2\,,
\ea
and the $(t,x)$ null vectors corresponding to this metric are
\begin{equation}
\label{eq:null}
v_1^\mu=\left(\begin{array}{c}
1\\
-1\\
0\\
0\\
\end{array}\right) \;\;\;\;\;\; v_2^\mu=\left(\begin{array}{c}
1+f^{\prime\prime}\\
1-f^{\prime\prime}\\
0\\
0\\
\end{array}\right)\,,
\end{equation}
so that  fluctuations are superluminal when $f^{\prime\prime}(x+t)<0$.

%Ref \cite{Evslin:2011vh} suggested that if four plane wave solutions were to converge together to form a square, it might be possible for a fluctuation to traverse them in such a way as to form a CTC. However when propagating from one side of the square to another it is found that $f^{\prime\prime\prime}$ diverges, and so such solutions are outside the regime of validity of the effective field theory, \cite{Evslin:2011vh}\footnote{We emphasize however, that considering the self-interactions to be classically large, does not necessarily break the regime of validity of the EFT, as is explained at the beginning of this section. However that fact that $f'''$ diverges, {\it does} break the EFT and such a configuration cannot be investigated within the framework of the theory at hand.}.

We will build on the principles of the proposal of Ref.~\cite{Evslin:2011vh} to construct a CTC for the Galileon. We consider a plane wave that travels in a loop along a periodic coordinate. Taking the $x$ coordinate to be periodic we must ensure that the function $f(x+t)$ is periodic under the shift $x\rightarrow x+L$.
Just as the cylindrical CTC discussed in section \ref{sec:GR} may be viewed as a special limit of a more localized CTC such as in spinning cosmic string, we can view the present example as an approximation to a Galileon plane wave traveling on say the trapped photon surface outside a black hole, or some equivalent construction where the periodic direction $x$ is replaced by an angular direction (the black hole example is not an ideal set up as in a Schwarzschild spacetime closed orbits for relativistic particles are unstable).

\subsection{Relevance of Two Metrics}

It is worth pointing out that even though the metric \eqref{eq:eff_metric} allows for superluminal modes,
the metric itself is still a compactified version of Minkowski spacetime. Working in terms of $u=t+x$ and $v=t-x$, the effective metric is simply
\ba
G\mn \d x^\mu \d x^\nu=-\d u \d v + f^{\prime\prime}(u) \d u^2+\d y^2 +\d z^2\,.
\ea
We can then make the change of coordinates,
\ba
\label{eq:coc}
V=v-f'(u)\,,
\ea
such that the effective metric is clearly that of flat cylinder
\ba
\label{eq:flatMet}
G\mn \d x^\mu \d x^\nu=-\d u \d V +\d y^2 +\d z^2\,.
\ea
In the change of coordinates \eqref{eq:coc}, notice that since $f^{\prime}(u)$ is periodic, $V$ satisfies the same identification as $v$, $V\equiv V-L$ along with $u \equiv u+L$. If we were for instance in a different situation where $f^{\prime\prime}$ had a zero mode, say $f^{\prime\prime}=\alpha=$constant, then the change of coordinates $v\to V=v-\alpha\, u$ would have required us to identify $V$ with $V-(1+\alpha)L$. The resulting flat spacetime would then have been cylindrical in a spacetime direction rather than just a purely spacelike direction.

The fact that the resulting effective metric \eqref{eq:flatMet} is simply that of a flat cylinder, tells us that by means of this metric alone, one would never be able to produce a CTC. However this is  a different situation to GR, because \eqref{eq:eff_metric} is not the metric seen by all the degrees of freedom living in that geometry, but only by Galileons. Other fields which are coupled in a standard way to gravity, will see the flat background metric $g\mn = \eta\mn$. Even though one could never produce a CTC if information was propagating on the causal light cone of either $g\mn$ or $G\mn$ alone, one can create CTCs by combining both metrics. To achieve this it is sufficient to propagate information by means of Galileon fluctuations along the null direction provided by $v_2$ in \eqref{eq:null} for a while, before transferring the information to another particle that propagates on null or timelike geodesics with respect to $g\mn$. It is these `bimetric CTCs' that we shall be concerned with in the following.

\subsubsection*{Does a Common Cauchy Surface Exist?}
We expect CTCs to form if it is not possible to foliate in the geometries described by $G\mn$ and $g\mn=\eta\mn$ via a common family of Cauchy surfaces.
To check if this is possible, suppose we perform the following coordinate transformation
\begin{equation}
t=H(x,y,z,\tau)\,,
\end{equation}
where $H$ is an arbitrary function which is periodic under $x\rightarrow x+L$. We would like to use the surfaces $\tau$= constant as Cauchy surfaces.  For these to describe a suitable Cauchy surface they must be space-like with respect to both the background metric $\eta\mn$ and the metric $G_{\mu\nu}$ felt by Galileon fluctuations.

For simplicity we may look at the induced geometry on surfaces of constant $y$ and $z$. A necessary condition for the surface to be spacelike with respect to $\eta\mn$ is if
\begin{eqnarray}
\eta\mn \d x^\mu \d x^\nu|_{t=H(x,y,z,\tau) ; \d y=\d z=\d\tau=0}=(1-(\partial_x H)^{2})\d x^2
>0
\end{eqnarray}
which requires
\begin{equation}
(\partial_x H)^2< 1\,.
\label{eq:spacelike1}
\end{equation}
Similarly, the surface is space-like with respect to $G_{\mu\nu}$ if
\ba
G\mn \d x^\mu \d x^\nu|_{t=H(x,y,z,\tau) ; \d y=\d z=\d\tau=0}  >0
\ea
which implies
\ba
\(1-(\partial_x H)^2+f^{\prime\prime}(H+x)(1+(\partial_x H)^2\)\d x^2>0\,,
\ea
or equivalently,
\begin{equation}
(1+\partial_x H)\Big(1-\partial_x H+f^{\prime\prime}(H+x)(1+\partial_x H)\Big)> 0\,.
\label{eq:spacelike2}
\end{equation}
Combining equations (\ref{eq:spacelike1}) and (\ref{eq:spacelike2}) we find
\begin{equation}
\frac{1-\p_x H(x,y,z,\tau)}{1+\p_xH(x,y,z,\tau)}+f^{\prime\prime}(H(x,y,z,\tau)+x) > 0\,.
\end{equation}
Since $H(x,y,z,\tau)$ is periodic in $x$, there must exist some $x_0$ for any time $\tau$ at which $\p_x H(x,y,z,\tau)|_{x=x_0}=0$.  Thus a common Cauchy surface can only exist if, at this point
\begin{equation}
1+f^{\prime\prime}(x_0+H(x_0,y,z,\tau))> 0\,.
\end{equation}
Since this relation must hold for all $\tau$, a Cauchy surface common to both metrics exists only if $f^{\prime\prime}(u)> -1$ for all $u$.  Conversely if there exists a $u$ for which $f^{\prime\prime}(u)\leq -1$ then there is no possible choice of surface which is spatial with respect to both metrics for all times $\tau$. As we will see in what follows, CTCs may only form if there exists a point $u$ for which $f''(u)\leq -1$ which implies that CTCs may only form when both metrics share no common Cauchy surfaces as expected.

\subsection{Bimetric CTCs}

\label{sec:CTCs}

To see explicitly how the CTC forms let us first find the evolution of Galileon fluctuations.
Geodesics moving in the $+x$ direction in the metric $G_{\mu\nu}$ obey
\begin{eqnarray}
t&=&x+\int_0^{t+x}f^{\prime\prime}(\tilde u)\;\d \tilde u\\
&=&x+f^{\prime}(t+x)\,,
\end{eqnarray}
where without loss of generality we have chosen to set $f^{\prime}(0)=0$.
If $T$ is the time taken to travel from $x=0$ to $x=L$, along the ``right-mover" null geodesic, then
\begin{equation}
T=L+f^{\prime}(T)\,.
\label{eq:galCTC}
\end{equation}
It is therefore clear that it is not possible to construct a curve that returns to the point $x=L$ within an elapsed time $T=0$. However one can choose the background configuration appropriately (\ie by choosing the periodic function $f(u)$) such that there can be solutions to \eqref{eq:galCTC} with negative $T$. Such solutions are not yet  CTCs, but  can be used as the basis of constructing bimetric CTCs in which information propagates on both metrics. 

One may consider the configuration where information is carried by Galileon fluctuations starting from the origin $\mathcal{O} (x=0, t=0)$ and going to a point $\mathcal{A}$ with spacetime coordinates $x=L\equiv 0$ and $t=-|T|$, with $|T|<L$ satisfying
\ba
\label{eq:boundf'}
f'(-|T|)=-(L+|T|)<-L\,.
\ea
For this condition to be satisfied, there must be at least one point in the interval $0<u<L-|T|$ where $f''(u)<-1$.  However since $f$ is periodic, there must also be at least one point in the interval $-|T|<u<0$ where $f''(u)>1$, and we can then infer that if one were to continue propagating information with the effective metric $G\mn$, the origin point $\mathcal{O}$ would not lie within the future light cone of the point $\mathcal{A}$. However if one considered interactions at the point $\mathcal{A}$ such that from that point on, information was instead carried by a standard massive particle propagating in the metric $g\mn=\eta\mn$, this particle could then remain stationary in space but travel forwards in time to reach $x=0$ at time $t=0$. Then we have created a CTC for information, even if no single particle traverses the entire curve. 
\vspace{5pt}

\subsection{Stability}

We have established that the background plane wave solution will admit bimetric CTCs - CTCs of information - provided we allow for interactions between the Galileon and another particle. These arise whenever there exists some region for which $f''(u)<-1$. Let us now analyze the stability of this configuration. We shall see that these configurations are unstable, and that the backreaction becomes arbitrarily large in precisely the same sense as in the pure GR example discussed in section~\ref{sec:GR}.

The equation of motion for Galileon fluctuations $\phi$ about the background $\pi=\pi_0(x^\mu)$ is
\begin{equation}
-4 \p_u \p_v \phi-4f''(u)\p_v^2 \phi+\p_y^2 \phi+\p_z^2 \phi=0
\end{equation}
whose solutions can be represented as superpositions of 
\begin{equation}
\phi(u,v,y,z)=e^{i[k_y y+k_z z-\kappa_n v]}g(u)\phi_0\,,
\end{equation}
where $\phi_0$ is constant and the function $g(u)$ has the form
\begin{equation}
g(u)=\exp\left\{-i\left[\frac{k^2}{4 \kappa_n}u -\kappa_n\, f^{\prime}(u)\right]\right\}\,.
\end{equation}
As before $k^2=k_y^2+k_z^2$ is the total momentum along the transverse directions.
Periodicity of $\phi$ is imposed by the following condition
\ba
-\kappa_n+\frac{k^2}{4\kappa_n}=\frac{2\pi n}{L} \hspace{20pt} {\rm for }\hspace{10pt} n \in \mathbb{Z}\,,
\ea
the periodicity of $f$ means that this condition is independent of the form of $f$. The Galileon field configuration is therefore apparently stable even when reaching the threshold $f''<-1$. This should actually come as no surprise since the effective metric $G\mn$ is nothing but Minkowski space in a different system of coordinates. Thus we do not expect any instability for fluctuations living on top of this geometry if these fluctuations are decoupled from the other metric. However as we have seen,  the metric $G\mn$ alone is not sufficient to produce CTCs. To build a closed time-like curve it was necessary for the Galileon to interact with a massive species which propagates according to the causal structure of the other background Minkowski metric $\eta\mn$.  We proceed to analyze whether the coupled system is stable when the background Galileon solution allows for the formation of closed time-like curves.

\subsubsection*{Adding a Coupling to Another Field}

In what follows, we therefore consider two different scalar fields. The first one $\phi$ corresponds to the Galileon fluctuation  $\pi=\pi_0(x+t)+\phi(x^\mu)$ where $\pi_0$ is given by \eqref{eq:pi0}, so that $\phi$ lives in the effective metric $G\mn$. The second field $\chi$ has a mass $m_\chi$ and lives on the background Minkowski metric $g\mn=\eta\mn$. If these two fields interact, then generically when expanding around the background configuration $\pi=\pi_0$ an effective coupling will be generated already at quadratic order.
To give an example, suppose that the original fields interact via the rather natural Galileon invariant term $-\frac{1}{2}\frac{g}{\Lambda^4}\chi \((\partial_{\mu} \partial_{\nu} \pi)^2-(\Box \pi)^2\)$. Expanding around the background $\pi=\pi_0+\phi(x^\mu)$ we will generate already in the free theory a term of the form
\be
\L_{\rm int}=-\chi\, \hat{\mu}\, \phi\,,
\ee
where in this case the operator $\hat{\mu} \, \phi = \frac{g}{\Lambda^4} \(\partial^{\mu} \partial^{\nu} \pi_0 \partial_{\mu} \partial_{\nu} \phi-\Box \pi_0 \Box \phi\) $. This operator plays the role of an effective kinetic coupling between two fields and more generally, to preserve the Galileon symmetry it must contain at least two derivatives acting on $\phi$. For the following argument the precise form of this coupling will not matter, only that some coupling exists already at quadratic order. Note that couplings like this will also be generated by quantum corrections even if not already present at tree level. We intentionally consider a quadratic coupling such that both fields interact already within linearized perturbed theory. If we were to consider only a higher order coupling, one would need to work to higher order to see the interaction arising as well as the associated instability.

The real contribution of the interaction is to regulate the square root that will appear in the expression for one of the field momenta in \eqref{eq:sr} and so to impose a definite notion of positive and negative frequency modes. This regularization occurs independently of the exact interaction we consider. To see this, we begin with the action to quadratic order
\begin{equation}
\mathcal{L}=-\frac12 \(G^{-1}\)^{\mu\nu}\p_\mu \phi\p_\nu \phi -\frac 12 \eta^{\mu\nu}\p_\mu \chi\p_\nu \chi-\frac 12 m_\chi^2\chi^2-\chi\, \hat{\mu}\, \phi\,,
\end{equation}
so that the coupled equations of motion are
\begin{eqnarray}
-4\partial_u\partial_v\phi -4 f^{\prime\prime}\partial_v^2\phi+\partial_y^2\phi+\partial_z^2\phi&=& \hat{\mu}^{\dagger} \chi\\
-4\partial_u\partial_v\chi+\partial_y^2\chi+\partial_z^2\chi-m_{\chi}^2\chi &=& \hat{\mu} \phi\,.
\end{eqnarray}
where $\mu^{\dagger}$ is the adjoint operator to $\mu$: $\int \d^4 x \chi \hat{\mu}\phi=\int \d^4 x \phi \hat{\mu}^{\dagger}\chi$.
If the background is translation invariant in the $v,y,z$ directions then on Fourier transforming, the potentially derivative coupling encoded by the operator $\hat{\mu}$ is replaced with a function $\mu(\kappa,\vk,u)$. Here again, we use the notation where $\kappa$ is the momentum in the $v$ direction, $k_y$ and $k_z$ are the momenta in the $y$ and $z$ directions and where $\vk=\{k_y,k_z\}$ and $\vy=\{y,z\}$, $k^2=k_y^2+k_z^2$.
Again the form of this function will not be crucial to the following argument.
The equations of motion become
\begin{equation}
-4i\kappa\,\partial_u\!\left(\begin{array}{c}
\phi\\
\chi
\end{array}\right)+\left(\begin{array}{cc}
-4\kappa^2f^{\prime\prime}+k^2 &\  \mu^{\dagger}(\kappa,\vk,u) \\
\mu(\kappa,\vk,u) & k^2+m_{\chi}^2
\end{array}\right)\left(\begin{array}{c}
\phi\\
\chi
\end{array}\right) =0\,.
\end{equation}
Although solving these equations exactly is extremely difficult, it is always straightforward to solve them in the WKB regime, \ie for sufficiently high momenta $k_u$ along the $u$ direction such that the variation of $k_u$ is small, namely
\ba
\frac{\d k_u}{\d u}\ll k_u^2\,.
\ea
Then the standard WKB solution is a good one
\ba
\phi &\sim& \bar \phi \, e^{i\int k_u \d u-i \kappa\, v+i \vk . \vy } \\
\chi &\sim& \bar \chi \, e^{i\int k_u \d u-i \kappa\, v+i \vk . \vy }\,,
\ea
where $\bar \phi$ and $ \bar \chi$ are slowly varying prefactors.
Working within that approximation, the system of equations has a solution if
\begin{equation}
\label{det}
{\rm det}\left(\begin{array}{cc}
-4\kappa^2f^{\prime\prime}+k^2+4\kappa k_u & \mu^{\dagger} \\
\mu & \ k^2+m_{\chi}^2+4\kappa k_u
\end{array}\right) =0\,.
\end{equation}
This determines the two possible $u$-modes
\ba
\label{eq:sr}
k_u^{(\sigma)}=-\frac{2k^2+m^2_\chi-4 f'' \kappa^2 + \sigma \sqrt{(4 f'' \kappa^2 +m_\chi^2)^2+4 \mu \mu^{\dagger}}}{8\kappa}\,.
\ea
where $\sigma = \pm 1$.
To get a sense of what this implies, consider modes of sufficiently large momenta $\kappa$ such that both the interacting and mass terms maybe be neglected, $\mu, m_\chi^2 \ll k^2$. Then we obtain
\ba
k_u^{(\sigma)}=-\frac 14 \frac{k^2}{\kappa}+\frac 12  \kappa \(f''- \sigma |f''|\)\,.
\ea
Comparing this with the result we would get in the absence of an interaction we see  that there is a crucial difference, even when taking the limit $\mu \to 0$ (or the interaction coupling $g\to0$). The momentum along the $u$ direction now not only depends on the periodic function $f''(u)$ but also on its absolute value. In the single field case, the choice of positive versus negative frequency modes is ambiguous because of the fact that the effective frequency depends on $f''$, and can change sign as we traverse the $x$ direction.
However as soon as the field $\phi$ interacts with $\chi$, the causal structure of $\chi$ is imposed upon $\phi$. The smallest interaction requires us to us to fix the notion of positive and negative frequency mode for $\phi$ once and for all independently of the sign of $f''$. As a result, the absolute value of $f''$ should then be taken into account which significantly changes the behaviour of the modes. 
At high momenta, the quantity that can be identified as the frequency or energy of each the mode is
\be
E^{(\sigma)} = \kappa-k_u^{(\sigma)}=\kappa +\frac 14 \frac{k^2}{\kappa}-\frac 12  \kappa\(f'' - \sigma |f''|\)\,.
\ee
This should be compared with the form of the energies in the absence of interactions
\be
E_{ \, \rm non-interacting}^{(\sigma)}=\kappa +\frac 14 \frac{k^2}{\kappa}-\frac 12  \kappa\(f'' - \sigma f''\)\,.
\ee
Thus we see that the limit $\mu \rightarrow 0$ of the vacuum for the interacting system, is not equivalent to the tensor product of the vacua for the non-interacting systems,
\ba
&& \lim_{\mu \rightarrow 0}E^{(\sigma)} \neq E_{ \, \rm non-interacting}^{(\sigma)} \nn \\
&&\lim_{\mu \rightarrow 0} |\text{interacting vacuum} \rangle_{\phi \, \cup \, \chi } \neq |0 \rangle_{\phi} \otimes |0 \rangle_{\chi} .
\ea
%When the modes are decoupled we may choose to align the sign choice from the square root, with the sign of $f''$ thus replacing $\pm |f''|$ with either $+ f''$ or $-f ''$. With this choice, when we impose the quantization condition $f''(u)$ will drop out. On the other hand, in the interacting case we must deal with the absolute value.
\subsubsection*{Quantization Conditions}

The energy for the fluctuations become quantized because of the periodicity conditions $\phi(u+L,v-L)=\phi(u,v)$, $\chi(u+L,v-L)=\chi(u,v)$, which determine the quantization condition for $\kappa$:
\ba
\int_0^L k_u^{(\sigma)} \d u + \kappa_n L=2\pi n\, \hspace{20pt}{\rm with }\hspace{10pt}n \in \mathbb{Z}\,.
\ea
This is equivalent to the statement
\ba
\(1 - \frac{ \sigma}{2L}\int_0^L |f''(u)|\d u\)\kappa_n^2-\frac{2\pi n}{L}\kappa_n-\frac 14 k^2=0\,,
\ea
and is easy to solve giving
\be
\kappa_{n,\varsigma}^{(\sigma)}  = \frac{1}{2\(1 - \sigma \A\) } \left( \frac{2\pi n}{L} +\varsigma \sqrt{\(\frac{2\pi n}{L} \)^2+k^2 \(1 - \sigma \A \)}\right)\,,
\ee
where we have defined the positive constant,
\ba
\mathcal{A}= \frac{1}{2L}\int_0^L |f''(u)|\d u>0\,.
\ea
The choice of $\varsigma=\pm 1$ in the expression for $\kappa_n$ is an independent sign choice from the one denoted by $\sigma$. Choosing the correct combinations of signs is important in determining the correct split into positive and negative frequency modes in order to quantize the system. The choice that corresponds to the positive frequency/energy modes is $\varsigma=+1$.
This equation admits a real solution for both sets of modes (\ie both $\sigma = \pm 1$ ) only if the background satisfies the condition
\ba
\label{eq:bound}
\mathcal{A}= \frac{1}{2L}\int_0^L |f''(u)|\d u<1\,.
\ea
Remarkably this condition is violated as soon as we enter a regime where CTCs may form, as is shown explicitly in the Appendix. Although not immediately obvious, this result follows from the periodicity of the function $f(u)$. Therefore as soon as we are in such a regime where CTCs may form, modes with large enough momentum $k$ have complex $\kappa$'s and in turn complex energies. The solution is therefore unstable with an arbitrarily small time scale set by the momentum $k$. This is similar to a ghost-like instability. If one were to start with an initial situation where \eqref{eq:bound} is satisfied and adiabatically increase $f''(u)$ so as to break that bound, the momentum $\kappa$ would diverge independently of the transverse momentum $k$. This is the indicator of strong coupling, as we shall see in section~\ref{sec:strongcoupling}. 

\subsubsection*{Eigenstates}

The coupling of the two modes means that the eigenstates of the system are a mixture of both $\phi$ and $\chi$ modes. The eigenstates are determined by solving
\begin{equation}
\left(\begin{array}{cc}
-4\kappa^2f^{\prime\prime}+k^2+4\kappa k_u^{(\sigma)} \ & \mu^{\dagger} \\
\mu &\ k^2+m_{\chi}^2+4\kappa k_u^{(\sigma)}
\end{array}\right)\left(\begin{array}{c}
\phi\\
\chi
\end{array}\right)  =0\,.
\end{equation}
Using the solutions for $k_u^{(\sigma)}$ we find that the relevant eigenstates of this system are
\begin{equation}
v^{(\sigma)}=\left(\begin{array}{c}
2\kappa^2f^{\prime\prime}+\frac{1}{2}m_{\chi}^2-\sigma \sqrt{(2  \kappa^2 f'' +\frac{1}{2}m_\chi^2)^2+ \mu \mu^{\dagger}}\\
\mu
\end{array}\right)\,.
\end{equation}
To illustrate the behaviour of these eigenstates more clearly we treat the interaction terms $\mu$, $\mu^{\dagger}$ as small, then
\begin{equation}
v^{(\sigma)}=\left(\begin{array}{c}
2\kappa^2f^{\prime\prime}+\frac{1}{2}m_{\chi}^2-\sigma |2  \kappa^2 f'' +\frac{1}{2}m_\chi^2|\\
\mu
\end{array}\right)+\mathcal{O}(\mu\mu^{\dagger}).
\end{equation}
The behaviour of the eigenstates is clearly determined by the sign of $2  \kappa^2 f'' +\frac{1}{2}m_\chi^2$, if we are working at low momentum ($\kappa\ll m_{\chi}$) then this is dominated by the mass term which is always positive, and to first order in $\mu$
\begin{equation}
v^{(+1)}=\left(\begin{array}{c}
0\\
\mu
\end{array}\right),\;\;\;\;\;\;
v^{(-1)}=\left(\begin{array}{c}
m_{\chi}^2\\
\mu
\end{array}\right).
\end{equation}
At high momenta, to first order in $\mu$
\begin{equation}
v^{(\sigma)}=\left(\begin{array}{c}
2\kappa^2(f^{\prime\prime}-\sigma|f^{\prime\prime}|)\\
\mu
\end{array}\right).
\end{equation}
If the top component of this vector is non-zero then the bottom component can be taken to be zero in the limit $\mu \rightarrow 0$. However, when the top component is zero, the bottom component cannot be neglected and can be normalized to unity. 
Given this we see that when $f^{\prime\prime}>0$ the eigenstates have the same form as at low momenta but when $f^{\prime\prime}<0$, which is precisely when we have superluminality, the roles of the eigenstates switch. Thus we have:
 \begin{itemize}
 \item At low momenta, the light state $\sigma =-1$ gives the eigenstate which is mostly $\phi$, and inherits its superluminality,
 and the heavy state $\sigma=+1$ gives the eigenstate which is mostly $\chi$ and (sub)luminal.
 \item At high momenta, whether the eigenstate for a given choice of $\sigma$ is mostly (superluminal) $\phi$ or mostly ((sub)luminal) $\chi$ depends on the local sign of $f^{\prime\prime}$ and thus oscillates in time and in space.
 \end{itemize}

\subsection{Strong Coupling and Quantum Backreaction}

\label{sec:strongcoupling}

We have established that by means of two metrics, one can produce CTCs only if there exists a point for which $f^{\prime\prime}< -1$. However this is precisely when the configuration is unstable due to the perturbations having complex energies. To further solidify the unphysical nature of the CTCs we will now see that the EFT inevitably breaks down before the CTC has a chance to form.

To see this we need to quantize the coupled fluctuations of both fields. The free Lagrangian for perturbations is given by
\ba
\mathcal{L}_\phi&=&\frac 12 (1+f'')(\p_t \phi)^2-f'' \p_t \phi \p_x \phi-\frac 12 (1-f'')(\p_x \phi)^2-\frac 12(\p_y \phi)^2-\frac 12(\p_z \phi)^2 \nn \\
&&-\frac{1}{2} \eta^{\mu\nu}\partial_{\mu} \chi\partial_{\nu} \chi -\frac{1}{2}m_{\chi}^2 \chi^2 - \chi \, \hat{\mu}\,  \phi\,.
\ea
In order to compute the conjugate momenta it is necessary to specify the dependence of the coupling operator $\hat{\mu}$ on time derivatives. We shall assume for simplicity that it is independent of time derivatives so that it does not contribute to the conjugate momentum of either field. In this case the conjugate momentum associated to $\phi$ is
\begin{equation}
p_\phi=(1+f'')\p_t \phi-f^{\prime\prime} \partial_x\phi\,,
\end{equation}
whereas that for $\chi$ is conventional; $p_{\chi} = \p_t \chi$. Then the Hamiltonian is
\begin{equation}
\mathcal{H}=\frac{1}{2\left(1+f^{\prime\prime}\right)}\(p_\phi+\partial_x\phi\)^2+\frac{1}{2}p_{\chi}^2+\frac{1}{2}(\partial_y\phi)^2+\frac{1}{2}(\partial_z\phi)^2+\frac{1}{2} (\vec{\nabla} \chi)^2+\frac{1}{2}m_{\chi}^2 \chi^2 + \chi \hat{\mu} \phi \,.
\end{equation}
From a classical perspective it is clear from this that the Hamiltonian diverges when $f^{\prime\prime}$ tries to reach the threshold $f''=-1$, which is a necessary condition for constructing a CTC, provided we assume that $p_\phi+\partial_x\phi \neq 0$. However on the other hand, if we can arrange $p_\phi+\partial_x\phi$ to vanish when $f''$ hits $-1$ then no obvious problems arise. To determine if there really is a  problem we must calculate the quantum expectation value of the Hamiltonian.

To quantize the fields we must split them up into positive and negative energy modes. However since the fields are coupled each field will be composed of the two independent positive energy solutions corresponding to the two choices of the sign $\sigma$.
The light eigenstate (the one that is mostly $\phi$ - and gives pure $\phi$ fluctuations in the limit $m_\chi \rightarrow \infty$) corresponds to the choice $\sigma =-1$ so that we have
\ba
\kappa_n^{(-)} \equiv  \kappa_{n,\varsigma=+1}^{(\sigma=-1)}=\frac{1}{2\(1+ \A\)} \left( \frac{2\pi n}{L} + \sqrt{\(\frac{2\pi n}{L} \)^2+k^2 \(1+ \A \)}\right)\,,
\label{eq:kappaL}
\ea
and the energies are given by
\be
E^{(-)}\equiv E^{(\sigma=-1)}_{\varsigma=+1} = \kappa_{n}^{(-)} +\frac{2k^2+m^2_\chi-4 f'' \kappa_{n}^{(-)}{}^2 - \sqrt{(4 f'' \kappa_{n}^{(-)}{}^2 +m_\chi^2)^2+4 \mu \mu^{\dagger}}}{8 \kappa_{n}^{(-)}}.
\ee
On the other hand the massive eigenstate (mostly $\chi$) corresponds to  the choice $\sigma = +1$ so that
\be
\kappa_n^{(+)} \equiv  \kappa_{n,\varsigma=+1}^{(\sigma=+1)}=
 \frac{1}{2\(1- \A\)} \left( \frac{2\pi n}{M} +\sqrt{\(\frac{2\pi n}{L} \)^2+k^2 \(1- \A \)}\right)\,,
\label{eq:kappaM}
\ee
and the energies are given by
\be
E^{(+)}\equiv  E^{(\sigma=+1)}_{\varsigma=+1} =\kappa_n^{(+)} +\frac{2k^2+m^2_\chi-4 f'' \kappa_n^{(+)}{}^2 + \sqrt{(4 f'' \kappa_n^{(+)}{}^2 +m_\chi^2)^2+4 \mu \mu^{\dagger}}}{8\kappa_n^{(+)}}\,.
\ee
To confirm that  the correct choices of sign have been made in (\ref{eq:kappaL}) and (\ref{eq:kappaM}) we can take the limit $f''(u) \rightarrow 0 $ and $\mu \rightarrow 0$ in which case we get the familiar result for the energy of a massless field in lightcone coordinates
$E^{(-)} \rightarrow \kappa + k^2/4\kappa$
along with that for a massive field $E^{(+)} \rightarrow \kappa + (k^2+m_{\chi}^2)/4\kappa$.

It is apparent from the above expressions that it is the massive eigenstate $\sigma = +1$ and not the light one whose energies become complex in the region of the formation of the CTC. Thus it is the backreaction from the massive eigenstate which will imply strong coupling and prevent the formation of the CTC. This result is somewhat surprising since it is the light eigenstate which is superluminal at low energies. However, the backreaction comes from the high energy behaviour of the modes where both eigenstates oscillate in time between superluminal and (sub)luminal fluctuations.

The Klein-Gordan inner product for the normalization of the mode functions is the sum of the two contributions from each field. Only this combined sum is conserved in the interacting case. 
\ba
\langle \phi' \chi'| \phi \chi \rangle =&&- i \int_{t={\rm constant}}\hspace{-30pt} \d^3x \left[ {\phi'}^* \((1+f'') \partial_t  \phi - f'' \partial_x \phi \)-      \((1+f'') \partial_t  {\phi'}^* - f'' \partial_x {\phi'}^* \) \phi  \right] - \nn \\
&& i \int_{t={\rm constant}}\hspace{-30pt} \d^3x  \left[ {\chi'}^* \partial_t  \chi -  \partial_t  {\chi'}^* \chi  \right] \, .
\ea

\subsubsection*{Light Eigenstate Fluctuations $\sigma=-1$}

In the light eigenstate $\phi$ can be expressed as a superposition of modes of the form
\be
 u_{n,\vk}(x,\vy,t) =\bar \phi \, e^{-i\(\int E^{(-)}(u,n,\vk) \d u\)+i \kappa^{(-)}_{n}(n,\vk)\, (u-v)+i \vk . \vy}\,,
 \ee
as well as its complex conjugate, and similarly for $\chi$. Unfortunately, although providing a complete set, this set of modes is not orthogonal on the constant $t$ hypersurfaces.
To see this let us consider the interactions $\mu$ to be negligible and the low momenta regime where the eigenstate is mostly $\phi$. The nontrivial form of the momentum conjugate to $\phi$, $p_{\phi}=(1+f'') \partial_t  \phi - f'' \partial_x \phi$ implies that the Klein-Gordan inner product takes the form
\be
\langle u' | u \rangle = - i \int_{t={\rm constant}}\hspace{-30pt} \d^3x \left[ {u'}^* \((1+f'') \partial_t  u - f'' \partial_x u \)-      \((1+f'') \partial_t  {u'}^* - f'' \partial_x {u'}^* \) u   \right].
\ee
Substituting in modes with different values of $n$ we obtain
\ba
\langle n',\vk' | n, \vk \rangle =\frac{|\bar \phi|^2}{2}(2\pi)^2\delta^{(2)}(\vk'-\vk)
\int_0^L \d x\Bigg\{\left[\frac{k^2}{4}\left(\frac{1}{\kappa_n}+\frac{1}{ \kappa_{n'}}\right)+(1+f^{\prime\prime}(x))(\kappa_n + \kappa_{n'})\right]\nn \hspace{-20pt} \\
\times \exp\(2i (\kappa_{n}^{(-)}-\kappa_{n'}^{(-)})x-i \int^x_0 \d\bar{x} (E^{(-)}(\bar{x},n,\vk)-E^{(-)}(\bar{x},n',\vk))\)\Bigg\} \,,\hspace{20pt}
\ea
which in general does not vanish for $n \neq n'$ even though it does vanish for different $k$. Note that the $t$ dependence does drop out because  the spatial $x$ integral is around a closed loop, the starting point of the integral is irrelevant and so we can replace $x+t$ by $x$ within the integrand. This confirms that this is the correct conserved inner product. The fact that the modes are not orthogonal is not surprising because the modes we have defined were obtained by Fourier transforming in the $v$ direction, and so will be orthogonal when using a definition of inner product on a constant $u$ hypersurface but not on constant $t$ hypersurfaces. Unfortunately this makes the direct quantization of modes using the constant $t$ Cauchy surfaces rather complex (although by no means impossible - since the set is complete there is no problem in principle).

Fortunately it is relatively easy to bypass this problem by means of the following trick. Define the new coordinate
\be
V = v - \int^u \d \bar{u} \left( \frac{1}{2} f''(\bar u) + \frac{1}{2} |f''(\bar u)| \right)\,.
\label{eq:coordV}
\ee
Without the modulus sign this is the same transformation we used before to show that the plane wave geometry is equivalent to Minkowski spacetime Eq.~(\ref{eq:coc}).
Working at sufficiently high momenta that we can neglect the mass $m_{\chi}$ of the second field (or indeed simply assuming the second field is massless) then the energies of the light eigenstate in the limit $\mu \rightarrow 0$ take the form
\be
E^{(-)} \approx \kappa_{n}^{(-)} +\frac{k^2-2  \kappa_{n}^{(-)}{}^2(f'' + |f''|)}{4\kappa_{n}^{(-)}}.
\ee
In this limit it is easy to see that the mode functions simply become
\be
 u_{n,\vk}(x,\vy,t) = \bar \phi \exp\({-i \( \frac{k^2}{4 \kappa_{n}^{(-)}}\) u -i \kappa_{n}^{(-)} V+i \vk . \vy} \),
 \ee
 which are precisely the mode functions for a massless particle in Minkowski spacetime with metric
 \be
 \d s_{(-)}^2 = -\d u \d V +\d y^2 + \d z^2.
 \label{eq:dsL}
 \ee
That is because in this limit, eq.~(\ref{eq:dsL})  describes the correct effective metric $G^{(-)}_{\mu\nu}$ that fluctuations of the  light eigenstate see. In other words, the light eigenstate continues to behave effectively like a massless field on Minkowski spacetime. The fact that it is necessary to perform a non-analytic coordinate transformation to see this Minkowski metric, is capturing the fact that the effective metric is really a combination of the metric seen by $\phi$ and that seen by $\chi$. Both of these are individually Minkowski, but not in the same coordinates.

The coordinate redefinition in  eq.~(\ref{eq:coordV}) that we performed to obtain this effective metric is not analytic, but it is  once differentiable which is sufficient for what follows. It can always be considered  as the limit of an analytic transformation. To make this clear  we undo the coordinate transformation, then the metric that describes the propagation of  fluctuations  of the light eigenstate is
\be
\d s_{(-)}^2 = -\d u \d v +\frac{1}{2}(f''(u) + |f''(u)| )\d u^2+ \d y^2 + \d z^2.
\ee
We can choose to regulate the metric in a manner consistent with the $\mu\rightarrow 0$ limit under which is was obtained
\be
\d s_{(-)}^2 = -\d u \d v +\frac{1}{2}(f''(u) + \sqrt{f''(u)^2+ \epsilon^2 } )\d u^2+ \d y^2 + \d z^2.
\ee
This is still a plane-wave geometry, and  all curvature invariants vanish for this metric. Thus there is no problem taking the limit $\epsilon \rightarrow 0$.

In this effective Minkowski space that describes the fluctuations of $\phi$, we can define a new time coordinate $T=\frac{1}{2}(u+V)$ and a new space coordinate $X= \frac{1}{2} (u-V)$. The identification which defines the period of the compactified dimension  which in the original coordinates takes $x \rightarrow x + L$, with $t$ left unchanged, in the new coordinates takes  $u\rightarrow u+L$ simultaneously with
 \be
 V \rightarrow  V-L(1+\A).
 \ee
 Equivalently we can identify new time and space coordinates $T=(1/2)(u+V)$ and $X=(1/2)(u-V)$ in which we identify
 \ba
 X &\rightarrow& X  + L  +  \frac{1}{2} L \A,\\
  T &\rightarrow& T - \frac 12 L \A.
 \ea
 The interval $s$ that defines the proper length of the closed curve which becomes the CTC described in section \ref{sec:strong1} is in the present case
 \be
 s = L\,  \sqrt{1+\A}\,.
 \ee
As we can see $s$ never passes through zero. Thus (at one-loop order) the quantum fluctuations of the light eigenstate do not blow up as we approach the region of the formation of the CTC. This is consistent with the fact that the energies of the light eigenstates always remain real and finite for finite momenta $k$ and $n$.

\subsubsection*{Massive Eigenstate Fluctuations $\sigma=+1$}

In the massive eigenstate $\chi$ can be expressed as a superposition of modes of the form
\be
 v_{n,\vk}(x,\vy,t) = \bar \chi \, e^{-i\(\int E^{(+)}(u,n,\vk) \d u\)+i \kappa_{n}^{(+)}(n,\vk)\, (u-v)+i \vk . \vy}
 \ee
as well as its complex conjugate. In this case, in the limit $\mu \rightarrow 0$, at low momenta  the eigenstate is mostly $\chi$ and the Klein-Gordan inner product is standard
\be
\langle v' | v \rangle = - i \int_{t={\rm constant}} \d^3x \left[ {v'}^* \partial_t  v -  \partial_t  {v'}^* v  \right],
\ee
and so the different modes are orthogonal and it is possible to employ a standard quantization procedure.
Nevertheless at high momenta, the eigenstates oscillate between $\phi$ and $\chi$, as in the case of the light eigenstate, and so to quantize it easier to use the previous trick and work with new coordinates.  We define
\be
\tilde{V} = v - \int^u \d \bar{u} \left( \frac{1}{2} f''(\bar u) - \frac{1}{2} |f''(\bar u)| \right).
\ee
Working, again,  at sufficiently high momenta that we can neglect the mass $m_{\chi}$ of the second field then the energies of the $\chi$ field in the limit $\mu \rightarrow 0$ take the form
\be
E^{(+)} \approx \kappa_{n}^{(+)} +\frac{k^2-2  \kappa_{n}^{(+)}{}^2(f'' - |f''|)}{4 \kappa_{n}^{(+)}},
\ee
and in this limit the mode functions simply become
\be
 v_{n,\vk}(x,\vy,t) = \bar \chi \exp\({i \( \frac{k^2}{4 \kappa_{n}^{(+)}}\) u +i \kappa_{n}^{(+)} \tilde{V}+i \vk . \vy} \) ,
 \ee
 which are again precisely the mode functions for a particle in Minkowski spacetime with metric
 \be
 \d s_{(+)}^2 = -\d u \d \tilde{V} +\d y^2 + \d z^2 = -\d u \d v +\frac{1}{2}(f''(u) - |f''(u)|)\d u^2+ \d y^2 + \d z^2.
 \label{eq:dsM}
 \ee
In the present case, in the superluminal region $f''(u)<0$ we see that the modulus sign acts to double the $f''(u)$ contribution and so in the high momenta regime  the massive eigenstate is superluminal with respect to the background metric, even though it always travels along its own effectively Minkowski metric. The CTCs are formed from curves which are timelike with respect to one of these metrics, but not necessarily both at the same time.  They should  be viewed as curves built out of segments where each segment is defined on one or the other of the metrics. These disjoint CTCs describe the effective propagation between the two metrics, and the  effective metric in Equation (\ref{eq:dsM}) is the one that describes the propagation of that information.

Finally we can again define a new time coordinate $\tilde{T}=\frac{1}{2}(u+\tilde{V})$ and a new space coordinate $\tilde{X}= \frac{1}{2} (u-\tilde{V})$ such that the cylindrical identification $x \rightarrow x+L$ becomes
 \ba
 \tilde X &\rightarrow& \tilde X  + L  -  \frac{1}{2} L \A  \\
  \tilde T &\rightarrow& \tilde T +  \frac{1}{2} L \A \,.
 \ea
The proper length of the loop $s$ from section \ref{sec:strong1} is in this case
 \ba
 s = L\,   \sqrt{1-\A}\,,
 \ea
 which could have been derived from \eqref{otw_interval} with $L\to L  -  \frac{1}{2} L \A$ and $A\to \frac{\A}{2-\A}$\,.
 Thus we may immediately borrow the result of section \ref{sec:strong1} to say that when $s$ approaches zero, the two-point function of the massive eigenstate modes diverges as $1/s^2$ and the expectation value of the total Hamiltonian diverges as $1/s^4$ because of the contributions coming from the massive eigenstate.
 The condition for $s$ to remain positive is
 \be
 \A=\frac{1}{2L}\int_0^L \d u |f''(u)|<1\,.
 \ee
However as already mentioned below Equation (\ref{eq:bound}) and proven in the Appendix, this condition is violated whenever there exists at least one point for which $f'' < -1$, and this is precisely the condition for CTCs. Thus we can conclude that the expectation value of the two-point function and the expectation value of the Hamiltonian of the system diverge precisely at the point of formation of the CTC, namely when $s=0$. This is consistent with the fact that the energies of the massive eigenstates become infinite, at finite $k$ and $n$, at the onset of formation of the CTC.
Since, even in the absence of gravity, this implies an infinite amount of backreaction on the Galileon field, we conclude that the formation of the CTC lies outside of the regime of validity of the Galileon effective field theory.

The surprise result of this analysis is that ultimately it is the two-point function and stress-energy of the massive eigenstate (the one which is mostly $\chi$ at low momenta but is an oscillating mix of $\chi$ and $\phi$ at high momenta) that diverges at the onset of formation of the CTC. This is because  in the interacting theory, although for momentum below the scale $m_{\chi}$ only the light eigenstate is superluminal when $f''(u)<0$, for larger momenta above the scale $m_{\chi}$ the massive eigenstate inherits the superluminality in an oscillatory manner. This behavior could never have been anticipated by looking at the two fields in the absence of interactions.

%%%%%%%%%%%%%%%%%%%%%%%%%%%%%%%%%%%%%%%%%%%%%%%%%%%%%%%%%%%%%%%%%%%%%
%%%% Outlook
\section{Outlook}
\label{sec:Outlook}
The possibility of creating CTCs within the standard framework of General Relativity has been well established for more than half a century, beginning with Kurt G\"odel in 1949.
Allowing for this possibility would dramatically shake up the most fundamental principle of physics. Fortunately within the context of GR, Hawking's Chronology Protection Principle is widely believed to prevent any such configurations from appearing. %However in a world where neutrinos are accused of traveling superluminally, one may investigate the range of this Protection Principle beyond General Relativity.

It has been well established that is is possible to construct EFTs with Lorentz invariant Lagrangians that nevertheless allow for fluctuations whose effective metric allows for faster than light propagation. This occurs in Galileon models and in massive gravity models. In this paper, we have explored the possibility of constructing CTCs for Galileons. Depending on whether the Galileons are seen as their own fundamental degrees of freedom, or instead as the Goldstone mode of a massive graviton, the situations in which CTCs may form may change slightly, however the nature of the Chronology Protection Principle is always the same and is generically applicable. In massive gravity, CTCs may be constructed on a cylindrical spacetime which are identical to solutions found in GR. However, just like in GR, we find that these solutions are quantum mechanically unstable, and if one were to start with healthy initial conditions, one would need to cross an infinitely strongly coupled region where the EFT is no longer a valid description to achieve a configuration where CTCs may form.

More generally, even if the Galileon is considered as a fundamental degree of freedom in its own right, one can easily find configurations where no single particle may propagate along a CTC, however information could ride between different species of particles,  which feel different effective metrics, and hence create a CTC. In this situation CTCs may only form if particles with different effective metrics interact, a situation which is inevitable in a gravitational theory where all species of particles interact at least via gravity. Whilst these configurations may appear classically stable with respect to one metric, we show that as soon as these interactions are taken into account, the result changes dramatically and the configuration becomes unstable precisely when CTCs may form. Hence within the regime of validity of the theory at hand, CTCs are not realized.

The arguments presented in the paper, do not represent a complete no-go for the construction of CTCs, and neither, as yet, does the Hawking Chronology Protection Principle. However within a simple class of configurations, we have established that the creation of CTCs is inexorably tied to strong coupling issues, making such configurations unreliable using the description at hand. As with the Hawking Chronology Protection Principle, much needs to be done to prove the general validity of this argument. For instance we have not explicitly shown that it will apply in the case of a spatially localized CTC. Nevertheless, there are strong reasons to believe that this new type of chronology protection, applicable in the case of theories with more than one effective metric and superluminalities,  is as generic as that proposed by Hawking.

%%%%%%%%%%%%%%%%%%%%%%%%%%%%%%%%%%%%%%%%%%%%%%%%%%%%%%%%%%%%%%%%%%%%%
%%%% Acknowledgments
%\vspace{30pt}
\acknowledgments

We would like to thank Louis Leblond, David Seery and Mark Wyman for valuable remarks and conversations.
CB, CdR and LH are funded by the SNF, and CB is also funded by a University of Nottingham Anne McLaren Fellowship. AJT would like to thank the  Universit\'e de  Gen\`eve for hospitality whilst this work was being completed.

\appendix
\section{Bounds of $f''$}
In this Appendix, we show explicitly that as soon as the Galileon configuration can admit CTCs, then the bound \eqref{eq:bound} is violated and the solution becomes unstable with an arbitrarily fast decay rate.

As we have seen in section  \ref{sec:CTCs}, CTCs may form if the periodic function $f(u)$ is such that $f'(0)=0$ and there exists a point $-|T|$  for which
$f'(-|T|)=-(L+|T|)<-L$.
We start by denoting by $u_0$ the minimum of $f'(u)$. For simplicity we take $u_0$ in the interval $-L<u_0<0$. From the previous bound, we immediately infer that
\ba
\label{eq:bound2}
f'(u_0)< -L\,.
\ea
We now denote by $u_n$ the points $-L<u_0<u_1<\cdots<u_N< 0$ where $f''(u_n)=0$. Knowing that $f'$ is negative and minimal at $u_0$, \ie that $f''$ is positive in the interval $[u_0,u_1]$ and changes sign at each point $u_n$, we infer that
\ba
\label{eq:fbounds}
\int_{u_n}^{u_{n+1}}|f''(u)|\d u=(-1)^n\(f'(u_{n+1})-f'(u_n)\)>0\,.
\ea
Knowing this, it is easy to see that the total integral within the interval $[u_0,0]$ is already greater than $L$. Denoting $u_{N+1}=0$, the integral is
\ba
\int_{u_0}^0 |f''(u)|\d u=\sum_{n=0}^{N} |f''(u)|\d u=\sum_{n=0}^{N}(-1)^n(f'(u_{n+1})-f'(u_n))\,.
\ea
We now first assume that $N=2M$ is even, so that
\ba
\int_{u_0}^0 |f''(u)|\d u=-f'(u_0)+2\sum_{n=1}^M(f'(u_{2n-1})-f'(u_{2n}))+f'(0)>L\,.
\ea
Similarly, if $N=2M+1$ is odd,
\ba
\int_{u_0}^0 |f''(u)|\d u=-f'(u_0)+2\sum_{n=1}^M(f'(u_{2n-1})-f'(u_{2n}))+2f'(u_{N})-f'(0)>L\,,
\ea
where we used the fact that $f'(u_{N+1})=f'(0)=0$, and from \eqref{eq:fbounds}  we have $f'(u_{2n-1})-f'(u_{2n})>0$ and if $N$ is even, $f'(u_N)>f'(u_{N+1})=0$.

We could now repeat precisely the same argument within the interval $[0,L-u_0]$. If $\bar N=\bar M+1$ is the total number of zeros of the function $f''$ within a period, $f''(u_n)=0$ for all $n=1,\cdots, \bar N$, then the integral within a period is
\ba
\int_{u_0}^{L-u_0}|f''(u)|\d u=-2 f'(u_0)+2\sum_{n=1}^{\bar M}(f'(u_{2n-1})-f'(u_{2n}))+2 f'(u_{\bar N})>2 L\,,
\ea
where we have used the fact that since $u_0\equiv (L-u_0)$ is a minimum, $f''$ is negative in the interval $[u_N,L-u_0]$, so that
$\int_{u_N}^{L-u_0}|f''(u)|\d u=f'(u_N)-f'(L-u_0)=f'(u_N)>0$.

We can therefore infer that if the background configuration is in a regime which admits CTCs, that is if there is at least one point along the cylinder where $f'(u)<-L$, then this together with the periodicity condition automatically implies that
\ba
\int_0^L |f''(u)|\d u>2 L\,.
\ea
%\ba
%\left.\begin{array}{c}
%f(u)=f(u+L) \ \forall u \\
%\exists \ u_0 \ {\rm s.t.}\ f'(u_0)<-L
%\end{array}\right\} \Rightarrow
%\int_0^L |f''(u)|\d u>2 L\,.
%\ea

\vspace{5pt}

%%%%%%%%%%%%%%%%%%%%%%%%%%%%%%%%%%%%%%%%%%%%%%%%%%%%%%%%%%%%%%%%%%%%%
%%%% Bibliography

%\newpage

\end{document}